%% file: main.tex
\documentclass[submit]{aiaa-tc}

 \title{Cash Accumulation Strategy based on Optimal Replication of Random Claims with Ordinary Integrals}

 \author{
  Renko Siebols 
    \thanks{Student, Department of Mathematics and Statistics,  Curtin University, Bentley, WA.}
}

 \AIAApapernumber{YEAR-NUMBER}
 \AIAAconference{Conference Name, Date, and Location}
 \AIAAcopyright{\AIAAcopyrightD{YEAR}}

\usepackage{pgfplots}
\pgfplotsset{compat=newest}
\pgfplotsset{plot coordinates/math parser=false}
\newlength\figureheight
\newlength\figurewidth

\usepackage{hyperref}
\usepackage{amssymb}
\usepackage{amsmath}
\usepackage{subcaption}
\usepackage{ragged2e}
\usepackage{listings}
\usepackage{graphicx}
\usepackage{bm}
\usepackage{hanging}
\usepackage{color} 
\definecolor{mygreen}{RGB}{28,172,0} 
\definecolor{mylilas}{RGB}{170,55,241}

\begin{document}

\maketitle
\input{sections/abstract}
\input{sections/introduction}
\input{sections/method}

\input{sections/results}
\input{sections/discussion}

\input{sections/conclusions}

\input{references}

\end{document}

%% file: sections/abstract.tex
\begin{abstract}
This paper presents a numerical model to solve the problem of cash accumulation strategies for products with an unknown future price, like assets. Stock prices are modeled by a discretized Wiener Process, and by the means of ordinary integrals this Wiener Process will be exactly matched at a preset terminal time. Three applications of the model are presented: accumulating cash for a single asset, for set of different assets, and for a proportion of the excess achieved by a certain asset. Furthermore, an analysis of the efficiency of the model as function of different parameters is performed. 
\end{abstract}

%% file: sections/introduction.tex
\section{Introduction}
Cash accumulation strategies deal with the question on how much money should be deposited into an account to be able to buy something at a specific price some time in the future. When the future price of the desired buy is known, cash accumulation strategies are quite straightforward. However, it becomes more complex when the future price is unknown and due to unpredictable changes. For example, looking at financial markets, stock prices seem to experience some kind of unpredictable behaviour. There a multiple factors influencing the stock, such as market sentiment, speculations, and demand and supply. \\

Therefore, the challenge is to generate a cash accumulation strategy that will accumulate the exact amount of cash at a terminal time in the future needed to buy a specific stock, taking into account the fact that the stock price at this terminal time cannot be known any time before the terminal time. A common approach to this problem is to assume that the changes in stock prices are basically small random daily changes. Using this assumption stock prices can be modelled as a random walk, that is generated by, for instance, a Wiener Process, or a Brownian Motion. According to the Martingale Representation Theorem, stochastic integrals can replicate these random variables from a Wiener Process (Dokuchaev, 2013a). This way, a solution can be found such that the final value of the random walk can be matched exactly. Backward stochastic differential equations (BSDE)  can be used to find this solution. (Brunick \& Shreve, 2013). However, it is hard to find an explicit and unique solution for a BSDE, and if a solution can be found, it can take much computational effort (Huijskens, 2013). Therefore, it is desired to find a simpler solution to this problem. A solution using ordinary integrals is presented by Dokuchaev (2013a). Using ordinary integrals an exact solution can be obtained, but this solution is not unique. Therefore, an optimized solution is presented.  \\

The aim of this report is to present a MATLAB model that can implement the equations presented by Dockuchaev (2013a; 2013b), using the technique of optimal replication of random claims with ordinary integrals, applied to the problem of cash accumulation strategy. \\

The structure of this report is as follows. Section 2 will describe the method used for this problem, presenting the equations for the general problem and how these equations will be incorporated in the model for this application. Section 3 will present numerical experiments performed with the model by showing 3 example cash accumulation policy problems, and 2 experiments analyzing the efficiency of the model. Subsequently, section 4 will present discussion points that arise from analyzing the model. 

%% file: sections/method.tex
\section{Method}
This section will describe the method for generating the cash accumulation strategy model. First the equations for the general problem are presented in \autoref{ss:general}. Afterwards, in \autoref{ss:appl}, it is shown how these equations can be adapted for the purposes of a cash accumulation strategy problem and how these equations can be incorporated in the model. Subsequently, in \autoref{ss:multi} and \autoref{ss:excess}, 2 extensions to the model are presented: respectively adding multidimensionality, and the accumulation of a proportion of an excess that is achieved by an asset. 

\subsection{General Problem Statement}
\label{ss:general}
Suppose $f$ is a random vector defined by \autoref{eq:f} (Dokuchaev, 2013a). The random vector is defined as the sum of its expected value $\textbf{E}f$ and the integral of random changes generated by a $d$-dimensional Wiener Process $w(t)$. The coefficient $k_f(t)$  can influence these random changes, and it may depend on the history of the Wiener Process up until the current time $t$ (Dokuchaev, 2013a).

\begin{equation}
    \label{eq:f}
    f = \textbf{E}f + \int_{0}^{T}k_f(t)dw(t)
\end{equation}

The aim this problem is to find a solution function $x(t)$ that, starting from a certain point $x(0)=a$ with $a \in \mathbb{R}^n $, has the exact same value as the random vector at the terminal time $T$, so $x(T)= f$ (Dokuchaev, 2013a). Generally speaking, $x(t)$ can be described by the differential equation as can be seen in \autoref{eq:dxdt}, subject to the just before mentioned boundary conditions. In this equation, $A \in \mathbb{R}^{nxn} $ is a given matrix, and $b 
\in \mathbb{R}^{nxn}$ is a non degenerate matrix (Dokuchaev, 2013a). 
\begin{equation}
    \label{eq:dxdt}
    \frac{dx}{dt} = A x(t) + b u(t), \quad t \in (0,T)
\end{equation}

As the solution for $x(t)$ is not unique, an optimal solution should be found. For this optimal solution in an integral norm , the value for $\textbf{E} \int_{0}^{T} u(t)^\top \Gamma (t) u(t) dt$ should be minimized (Dokuchaev, 2013a). The parameter $\Gamma(t) = g(t)G(t)$ is related to the penalty for fast growing $u(t)$ as $t \rightarrow T$ (Dokuchaev, 2013a). For $g(t)$, the restrictions shown in \autoref{eq:res} and \autoref{eq:res2} apply over the entire time interval $(0,T]$, and $G(t) > 0$ has to be a symmetric positively defined matrix (Dokuchaev, 2013a).

\begin{equation}
    \label{eq:res}
    0 < g(t) \leq c(T-t)^{\alpha} 
\end{equation}

\begin{equation}
    \label{eq:res2}
    g(t)^{-1} \leq c(1+(T-t)^{- \alpha}
\end{equation}

Defining the following for parameters, the optimal solution for $u(t)$ can be expressed using \autoref{eq:optimal} (Dokuchaev, 2013a). In combination with \autoref{eq:dxdt} and the specified boundary conditions, $x(t)$ can be found (Dokuchaev, 2013a). 

\begin{equation*}
    \widehat{k}_{\mu}(t) = R(t)^{-1}k_{f}(t), \quad R(s) \triangleq  \int_{s}^{T} Q(t)dt, \quad  Q(t) = e^{A(T-t)} b \Gamma(t)^{-1}b^\top e^{A^\top (T-t)}
\end{equation*}

\begin{equation*}
    \widehat{\mu} = R(0)^{-1} (\textbf{E}f - e^{AT}a) + M(t) , \quad \text{with} \quad M(t)=\int_{0}^{t} \widehat{k}_{\mu}(t)dw(s)
\end{equation*}

\begin{equation}
    \label{eq:optimal}
    \widehat{u}(t) = \Gamma(t)^{-1}b^\top e^{A^\top (T-t)} \widehat{\mu}.
\end{equation}

\subsection{Application to Cash Accumulation Policy Model}
\label{ss:appl}
This general optimal solution can now be applied to create the desired cash accumulation strategy model. Imagine an investor wanting to buy an asset with stock price S(t) at a specified terminal time $T$. The stock price $S(t)$ is subject to random continuous changes $dS(t)$ that can be described by \autoref{eq:ds}, in which $a(t)$ can be described as a drift of the stock price, and $\sigma (t) dw(t)$ is the randomly changing part induced by the continuous Wiener Process $w(t)$ with $w(0)=0$, as could comparably be seen in \autoref{eq:f} (Dokuchaev, 2013a).

\begin{equation}
    \label{eq:ds}
    dS(t) = S(t) [a(t)dt +  \sigma (t) dw(t)]
\end{equation}

However, for modelling purposes, a discrete Wiener Process will be applied with time step $ \Delta t$, such that over the time interval $(0,T]$ $ N=\frac{T}{\Delta t} $ times steps will be present. In this discretization, the change in the random walk $dW$ is normally distributed, with mean $ 0$ and standard deviation  $\sqrt{ \Delta t}$ (Highman, 2013). The stock price can now be expressed using \autoref{eq:s}.

\begin{equation}
    \label{eq:s}
    S(t_k) = S(0) + S(t_{k-1})  \sum_{k=1}^{N} \big( a(t_{k-1}) \Delta t + \sigma(t_{k-1})dW(t_k) \big) 
\end{equation}

Consider the first and easiest case, in which there are no interest rates to consider. In this case, $A=0$, $b=1$, $n=1$, and $f=S(T)$. The function $u(t)$ describes the density of cash deposits and withdrawals, in units of dollars per unit time step. Therefore, the amount of cash deposited or withdrawn within one time step can be represented by $u(t) \Delta t $ (Dokuchaev, 2013b). The value for $a$ represents the starting amount of cash in the account, which may take both positive and negative values. A negative value for $a$ would indicate a debt at starting time $0$. With these assumptions, the function R(s) simplifies to \autoref{eq:r} and the optimal solution for $u(t)$ is found using \autoref{eq:u} (Dokuchaev, 2013b). Integrating $u(t)$ over the entire interval ensures that the integral matches the terminal value $f=S(T)$ exactly, as according to \autoref{eq:ut} (Dokuchaev, 2013b).

\begin{equation}
    \label{eq:r}
    R(s) \triangleq \int_{s}^{T} \Gamma (t) ^{-1}dt. 
\end{equation}

\begin{equation}
    \label{eq:u}
    u(t) = \Gamma (t) ^{-1} \left[ R(0)^{-1} (S(0) - a)+ \int_{0}^{t} R(s)^{-1} dS(t) \right]
\end{equation}

\begin{equation}
    \label{eq:ut}
    \int_{0}^{T} u(t)dt = f 
\end{equation}

In the next case, there are interest rates to consider, and for both loans and savings the interest rate is $r \geq 0$. For this case, $A=r$. Now, the functions for $Q(t)$ and $R(s)$ can be evaluated using \autoref{eq:qr} and \autoref{eq:rr} (Dokuchaev, 2013b). Subsequently, the cash flow density $u(t)$ can be evaluated using \autoref{eq:ur} (Dokuchaev, 2013b). To find the final amount of cash in the account at the terminal time $T$, the cash flows deposited at time $t$ should be accumulated using the accumulation factor $e^{r(T-t)}$. In accordance with \autoref{eq:utr}, the integral of the accumulated cash flows over the entire interval will match the terminal value $f=S(T)$ exactly (Dokuchaev, 2013b). 

\begin{equation}
    \label{eq:qr}
    Q(t)= e^{2r(T-t)}\Gamma (t) ^{-1}
\end{equation}

\begin{equation}
    \label{eq:rr}
    R(s) = \int_{s}^{T} Q(t)dt
\end{equation}

\begin{equation}
    \label{eq:ur}
    u(t) = \Gamma (t) ^{-1} \left[ R(0)^{-1} (S(0) - e^{rT}a)+ \int_{0}^{t} R(s)^{-1} dS(t) \right]
\end{equation}

\begin{equation}
    \label{eq:utr}
    \int_{0}^{T} e^{r(T-t)}u(t) = f
\end{equation}

This last case is the case that will be used for the model, as for $A=r=0$, all equations will be similar to the first case. First, for each time step $M(t_k)$ will be calculated using \autoref{eq:M}. Subsequently, $\mu(t_k)$ will be calculated using \autoref{eq:mu}. Afterwards, the cash flow density $u(t_k)$ will evaluated using \autoref{eq:utk}. Finally, the amount of accumulated cash in the account $x(t)$ can be evaluated using \autoref{eq:xtk}. In this equation, for each time step the cash  that is deposited or withdrawn, is accumulated to the end of the time step using the accumulation factor $e^{r \Delta t}$.

\begin{equation}
    \label{eq:M}
    M(t_k) = M(t_{k-1}) + R(t_k)^{-1}dW(t_k)
\end{equation}

\begin{equation}
    \label{eq:mu}
    \mu(t_k) = R(0)^{-1}(S(0)-e^{(rT)}a)+M(t_k)
\end{equation}

\begin{equation}
    \label{eq:utk}
    u(t_k) = \Gamma(t)^{-1}e^{r(T-t_k)}\mu(t_k)
\end{equation}

\begin{equation}
    \label{eq:xtk}
    x(t_k) = \big( x(t_{k-1}) + u(t_k) \Delta t \big) e^{r \Delta t}
\end{equation}

Plotting both $S(t)$ and $x(t)$ will give a visual representation of the development of the stock price $S(t)$ and the amount of cash $x(t)$ over time. Using the above equations, at terminal time $T$, $S(T)=x(T)$.

\subsection{Extension of Model: Accumulation Cash for Set of Different Assets}
\label{ss:multi}
The current model is suitable if an investor wants to buy one or a multiple of the same asset. Now, the model will be extended for the case that this investor wants to buy a package consisting of different assets at the terminal time. \\

Instead of all equations being scalar equations, for this situation, they will have to be vector equations. In \autoref{eq:sd}, \textbf{S$(t_k)$} now represents the vector in $\mathbb{R}^m$, where $m$ stands for the number of different assets, containing all individual stock prices at time step $t_k$. Similarly to the case in which only one stock was bought, every different stock will experience changes induced by an independent Wiener Process. Therefore, each entry of \textbf{dW}($t_k$) contains a random number from a normal distribution with mean 0 and standard deviation $\sqrt{\Delta t}$. This will ensure that each stock price changes independently, and that the changes in one stock price do not influence any other stock price.

\begin{equation}
    \label{eq:sd}
    \textbf{S$(t_k)$}= \textbf{S$(0)$} +\textbf{ S$(t_{k-1})$)}  \sum_{k=1}^{N} \big( \textbf{a$(t_{k-1})$} \Delta t + \bm{\sigma} \textbf{$(t_{k-1})$dW$(t_k)$} \big) 
\end{equation}

To solve this problem of buying multiple different assets, the model will now initiate an individual cash accumulation process for each different asset. Note that all deposits and withdrawals for each individual process are from the same bank account. Therefore, the interest rate $r$ will be the same for each process. \\

\autoref{eq:Md} and \autoref{eq:mud} are the vector equivalent versions of \autoref{eq:M} and \autoref{eq:mu}, with all vectors in $\mathbb{R}^m$. The vector $\textbf{a}$ represents the starting amount of cash that is available for each individual cash accumulation process. If a company has for instance $\$ 80 $ as a starting amount of cash, and wishes to buy 4 different assets, it will be assumed that $ \$ 20$ is available as a starting amount for each individual process.

\begin{equation}
    \label{eq:Md}
   \textbf{ M$(t_k)$} = \textbf{M$(t_{k-1})$} + R(t_k)^{-1}\textbf{dW$(t_k)$}
\end{equation}

\begin{equation}
    \label{eq:mud}
    \bm{\mu}(t_k) = R(0)^{-1}(\textbf{S$(0)$}-e^{(rT)}\textbf{a})+\textbf{M$(t_k)$}
\end{equation}

The vector function $\textbf{u}(t_k)$ now represents the cash flow density for each individual process, and is given by \autoref{eq:utkd}. Subsequently, the amount of cash that is accumulated in each process $\textbf{x}(t_k)$, can be calculated using \autoref{eq:xtkd}. These equations ensure that at the terminal time, each individual process exactly matches its corresponding stock price at the terminal time.

\begin{equation}
    \label{eq:utkd}
    \textbf{u}(t_k) = \Gamma(t)^{-1}e^{r(T-t_k)}\bm{\mu}(t_k)
\end{equation}

\begin{equation}
    \label{eq:xtkd}
    \textbf{x}(t_k) = \big(\textbf{x}(t_{k-1}) + \textbf{u}(t_k) \Delta t \big) e^{r \Delta t}
\end{equation}

However, since all cash for each process will be accumulated in the same bank account, it is more interesting to know what the total amount of cash is that has to be deposited or withdrawn at each time step.  This can be represented by $sum(\textbf{u}(t_k)) \Delta t$. The total amount of cash that is available in the account at each time step, can correspondingly be represented by $sum(\textbf{x}(t_k))$. The total cash in the account at the terminal time will now exactly match the sum of all different stock prices at the terminal time.

\subsection{Extension of Model: Accumulation of Proportion of Equity Excess}
\label{ss:excess}
Another extension of the model can be obtained when looking at the problem when the accumulated amount of cash has to be a certain proportion of the positive difference between the stock price and a certain strike price that is set at the start of the process. If there's no difference, or the difference turns out to be negative, the function will take value $0$. Therefore, the function $f$ will be given by \autoref{eq:fb} (Dokuchaev, 2013b) . In this, equation $c$ represents the proportion, and $K$ represents the preset strike price that will be used to calculate the achieved excess. An alternative expression for $f$, that is used in the model, is \autoref{eq:fh} (Dokuchaev, 2013b). \\

\begin{equation}
    \label{eq:fb}
    f = c \max (S(T)-K,0)
\end{equation}

\begin{equation}
    \label{eq:fh}
    f = H(S(0), 0) + \int_{0}^T \frac{\partial H}{\partial x}(S(t), t)dS(t).
\end{equation}

In this formulation, $H(S(t),t)$ can be calculated using the Black-Scholes formula for a call option (Dokuchaev, 2013b). Using this formula, an estimate can be calculated for an European call option. The expression for the Black-Scholes equation can be seen in \autoref{eq:bs} (Hull, 2012). In this equation, $N(\cdot)$ stands for the cumulative normal distribution. The variables $d_1$ and $d_2$ can be calculated using \autoref{eq:d1} and \autoref{eq:d2} respectively (Hull, 2012). Note that $\sigma$ in these equations should be constant, so $\sigma (t) = \sigma$ compared to \autoref{eq:ds} (Dokuchaev, 2013; Hull, 2012).

\begin{equation}
    \label{eq:bs}
    H(S(t),t) = N(d_1)S(t) - N(d_2)Ke^{-r(T-t)}
\end{equation}

\begin{equation}
    \label{eq:d1}
    d_1 = \frac{1}{\sigma \sqrt{T-t}} \left[ \ln \left( \frac{S(t)}{K}\right) + \left( r + \frac{\sigma^2}{2}\right)(T-t)\right]
\end{equation}

\begin{equation}
    \label{eq:d2}
    d_2 = d_1 - \sigma \sqrt{T-t}
\end{equation}

Comparing \autoref{eq:fh} with \autoref{eq:f}, it can be seen that the coefficient $k_f$ from \autoref{eq:f} can be calculated using \autoref{eq:kf}, again assuming that $\sigma$ is constant (Dokuchaev, 2013b). Subsequently, $\frac{\partial H}{\partial x}$ can be calculated using \autoref{eq:hx} (Haug, 2007). Whereas $k_f(t)$ could be any random constant in \autoref{eq:f}, it has to be calculated for each independent time step for \autoref{eq:fh}. Therefore, $k_f(t)$ can now be considered a function of time, that depends on the history of the random claim itself. The cash flow density for this problem can now be calculated using \autoref{eq:ubs} (Dokuchaev, 2013b).

\begin{equation}
    \label{eq:kf}
    k_f = \frac{\partial H}{\partial x} (S(t), t) \sigma dS(t)
\end{equation}

\begin{equation}
    \label{eq:hx}
    \frac{\partial H}{\partial x} (S(t), t) = N(d_1)
\end{equation}    
    
\begin{equation}
    \label{eq:ubs}
    u(t) = c \Gamma(t)^{-1} \left[ R(0)^{-1}H(S(0),0) + \int_{0}^{t} R(s)^{-1} \frac{\partial H}{\partial x} (S(t), t) \sigma(t) dS(t) \right]
\end{equation}

For the implementation of this application in the model, several equations have to be adapted slightly. Whereas in the in the previous example the stock prices could directly be used for the replication, now the function $f$ will have to be calculated first. Therefore, for each time step, the value for $f(t_k)$ will be calculated using \autoref{eq:fbl}. Subsequently, $M(t_k)$ and $\mu (t_k)$ will be calculated according to \autoref{eq:mbl} and \autoref{eq:mubl} respectively. The cash flow density $u(t_k)$ can now be represented by \autoref{eq:ubl}. 

\begin{equation}
    \label{eq:fbl}
    f(t_k) = f(t_{k-1}) + N(d1(t_k))\sigma dS(t_k)
\end{equation}

\begin{equation}
    \label{eq:mbl}
    M(t_k) = M(t_{k-1}) + R(t)^{-1}  N(d1(t_k))\sigma dS(t_k)
\end{equation}

\begin{equation}
    \label{eq:mubl}
    \mu(t_k) = R(0)^{-1}H(S(0),0) + M(t_k)
\end{equation}

\begin{equation}
    \label{eq:ubl}
    u(t_k) = c \Gamma(t)^{1}e^{A(T-t)}\mu(t_k)
\end{equation}

%% file: sections/results.tex
\section{Numerical Experiments}
This section will present the main results obtained from the implementation from the last section into the MATLAB model. Different experiments will be presented in order to show how the model behaves. In \autoref{sec:example}, 3 numerical examples will be presented, and in \autoref{sec:anal}, 2 experiments regarding the time performance of the model will be presented.

\subsection{Application of Model to Example Problems}
\label{sec:example}
\subsubsection{Accumulating Cash to Buy Asset after 1 Year}
For this first example, a company wanting to accumulate enough cash in order to be able to buy an asset after exactly 1 year is considered. The price of this asset at time 0 $S(0)$ is $\$ 150 $. The company currently has $\$ 50 $ available in the account. Given an interest rate $r$ of $12 \%$ for both loans and savings, the company wishes to automatically make a withdrawal or deposit once a day such that after 1 year, the company has exactly enough money in order to buy this asset. \\

For this rendering, the following parameters have entered into the program. The drift of the stock price $a(t)$ is assumed to be 0, and the coefficient $\sigma (t)$, that influences the random changes of the stock price is assumed to be 0.5. Note that these two parameters can have any value in $ \mathbb{R}^n$, or may be any function of time in order to match the stock price at the terminal time. Furthermore, $A=r=0.12$, $b=1 $, and $\Gamma(t)=1$. Since the company wants to update their account once a day for 1 year, the number of time steps $N$ is 365. \\

\begin{figure}
    \centering
    \input{tikz/rendering1.tex}
    \caption{Rendering of Model for Example 1}
    \label{fig:plot1}
\end{figure}
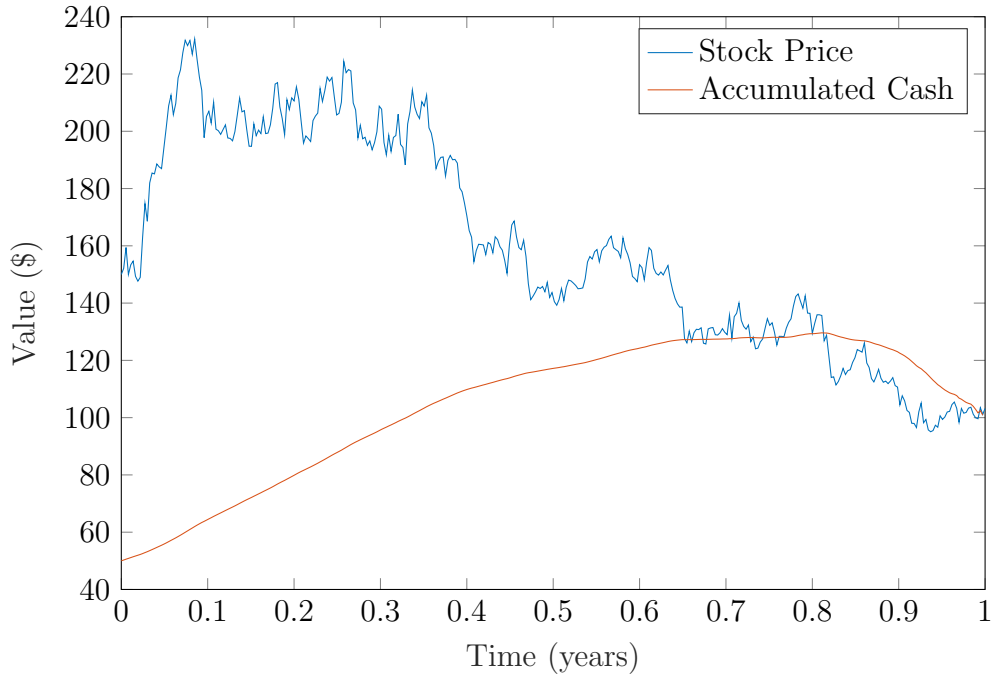

The rendering of the model for this example with the current parameters can be seen in \autoref{fig:plot1}. Both the development of the stock price and the amount of cash in the account can be seen. Looking at the plot, it can be noted that the changes in the stock price might be relatively big, but the changes in the amount of accumulated cash remain quite smoothly. Therefore, the amount of cash deposited or withdrawn is relatively small at each update. At the terminal time of 1 year, it can be seen that the amount of cash matches the stock price as desired. \\

If the company would have liked to buy a multiple of this asset, the model will have to be adapted slightly. If the starting stock price $S(0)$ as well as the observed changes $dS$ at each time step are multiplied by the desired number of assets, running the model will ensure that cash is accumulated to buy this number of assets at the terminal time of 1 year.

\subsubsection{Accumulating Cash to Buy Set of 2 Independent Stocks after 1 Year}
For the second example, now consider the same company wanting to buy a set of two different, independently behaving assets after exactly 1 year. At time 0, the first asset has a stock price $S(0)$ of $\$200$, whereas the second asset has a stock price $S(0) $ at time 0 of $\$400$. Currently, the company has a starting amount of cash of $\$40$ and wishes to update their account once every day, given that the interest rate $r$ is now $30 \%$.\\

The model will now be used to simultaneously accumulate cash to exactly buy both assets after one years. Similarly to the first example, $a(t) = 0$ and $\sigma (t)= 0.5$. The other parameters are $A=r=0.3$, $b =1$, and $\Gamma(t) = 1$. Updating the account once a day for 1 year leads to the number of time steps $N$ being equal to 365. \\

The model will work as follows. Each time step, both stock prices will change independently. Subsequently, the model will try to match both stock prices independently by withdrawing or depositing the right amount of cash. As the starting amount of cash is $\$ 40$, it will be assumed that the starting value for matching each independent stock price is $\$ 20$. So, by depositing the independent amounts of cash at each time step in the same account, this will ensure that there will be enough cash in the account after one year in order to be able to exactly buy buy both assets at the terminal time of 1 year. \\

\begin{figure}[h!]
    \centering
    \begin{subfigure}[t]{0.5\textwidth}
        \centering
        \input{tikz/22.tex}
        \caption{\centering Rendering of Independent Stock Price Behaviour for 2 Stocks}
        \label{fig:2stocks1}
    \end{subfigure}%
    ~ 
    \begin{subfigure}[t]{0.5\textwidth}
        \centering
        \input{tikz/21.tex}
        \caption{\centering {Rendering of Cash Accumulation to Buy Set of 2 Independent Stocks}}
        \label{fig:2stocks2}
    \end{subfigure}
    \caption{Rendering of Cash Accumulation to Buy Set of 2 Independent Stocks}
    \label{fig:2stocks}
\end{figure}
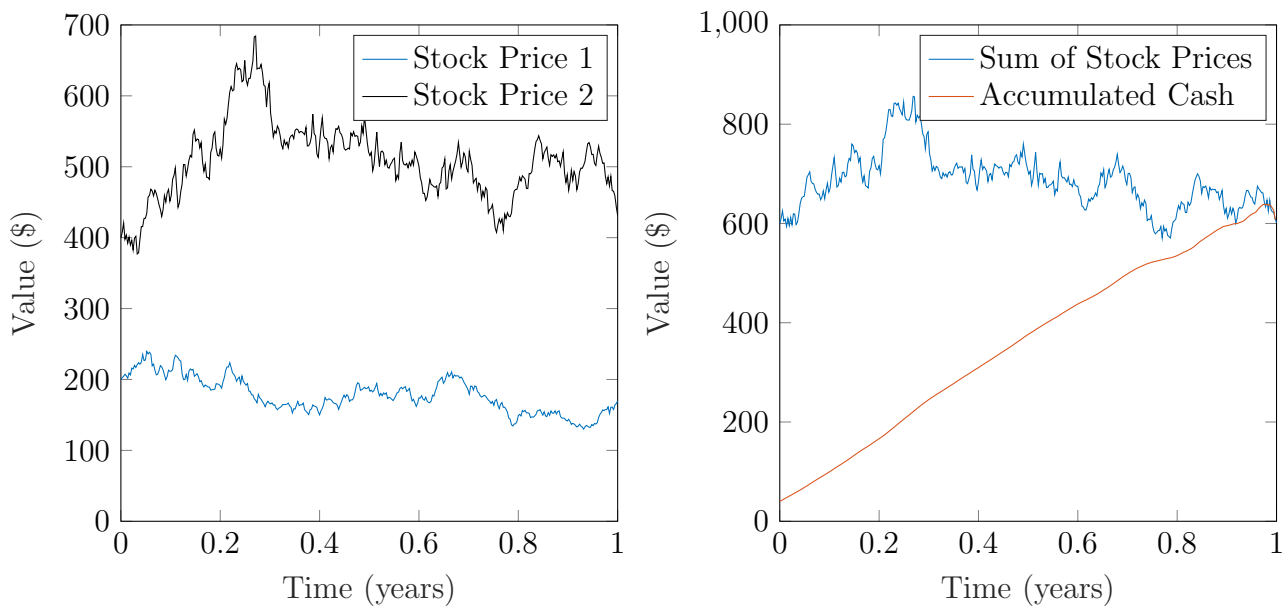

In \autoref{fig:2stocks}, two renderings for this example are presented. In \autoref{fig:2stocks1}, the behaviour of the stock prices of both assets can be found. As can be seen, the stock prices behave completely independently, and the changes of one stock price do not influence the changes the other stock price. Subsequently, in \autoref{fig:2stocks2}, the sum of the two independent stock prices is presented as well as the total amount of cash that is accumulated in the account. As can be seen, now the total amount of accumulated cash at the terminal time of 1 year, matches the sum of the two stock prices. \\

Similarly to the first example, the model should be changed slightly if the company wants to buy a multiple of each of the assets. As in the model $S(0)$ and $dS$ are now $2x1$ vectors, the inner product of these vectors and the $2x1$ vector containing the desired number of assets should be taken. Subsequently, running the program will ensure that the accumulated amount of cash in order to exactly buy the desired amount of each asset.

\subsubsection{Accumulating 50 \% of the Excess Achieved by Equity over 2 Years}

For the last case example, the accumulated cash has to be $50\%$ of the excess achieved by a certain equity. The starting price $S(0)$ is $\$ 75$ and the strike price $K$ is $\$ 30 $. Cash has to be accumulated in 2 years time, updating the account once every 2 days. The interest rate $r$ for this example is $3\%$.\\

For this example, $a(t)=0$, $\sigma (t)=0.3$. The other parameters are $A=r=0.03$, $b=1$, and $\Gamma (t)=1$. The proportion factor $c$ is 0.5. Updating the account once every 2 days for 2 years, leads to $N=365$.\\

\begin{figure}
    \centering
    \input{tikz/bls.tex}
    \caption{Rendering for Cash Accumulation of 50 \% of the Excess}
    \label{fig:bls}
\end{figure}
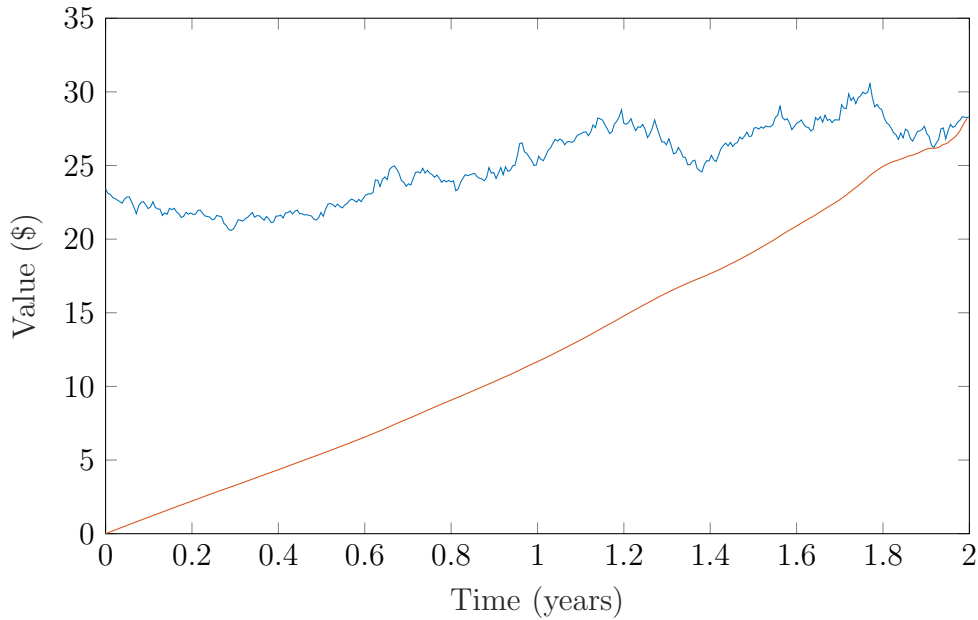

\autoref{fig:bls} shows the rendering of the model for this example. Instead the stock price being plotted as in the other examples, now the proportion of the current excess is plotted. As expected, the accumulated cash matches this excess at the terminal time of 2 years. 

\subsection{Model Performance Analysis}
\label{sec:anal}
This subsection will describe the influence of choosing different parameters on the overall performance of the model. Both the effect of changing the time step size and the number of dimensions are considered.

\subsubsection{Impact of Discretization Rate}
\begin{figure}[h!]
    
    \centering
    \begin{subfigure}[t]{0.5\textwidth}
        \centering
        \input{tikz/timestep1.tex}
        \caption{$\Delta t = 0.01$}
    \end{subfigure}%
    ~ 
    \begin{subfigure}[t]{0.5\textwidth}
        \centering
        \input{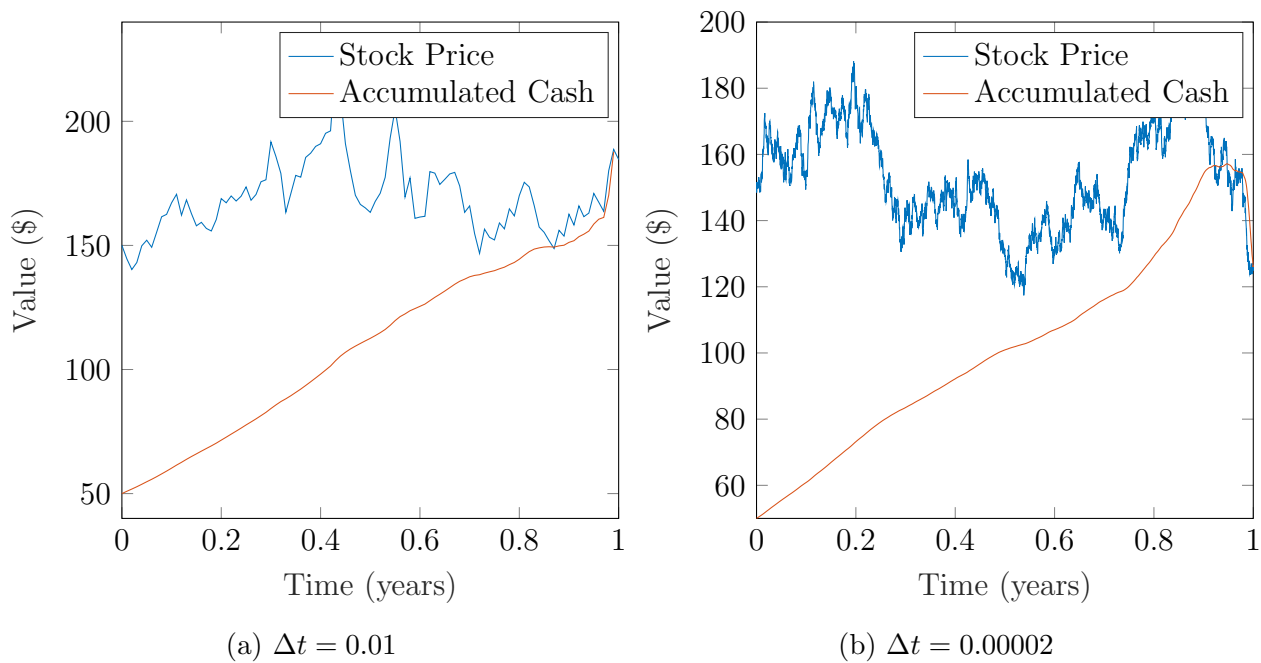}
        \caption{$\Delta t=0.00002$}
    \end{subfigure}
    \caption{Renderings for different values of $\Delta t$}
    \label{fig:timestep}
\end{figure}

First the impact of changing the time step size is considered. Changing the time step size changes the way the stock price behaves. As the changes in the random walks $dW$ are normally distributed with mean 0, and standard deviation $\sqrt{\Delta t}$, for smaller time steps $\Delta t$ the change at each time step is also expected to be smaller. However, as the time steps are smaller, this means that over the entire time interval, more time steps will have to be evaluated. Therefore, the absolute changes that will be experienced are comparable to the case with larger time steps. This can also be seen in \autoref{fig:timestep}, where two the model is run with two different time step sizes. 

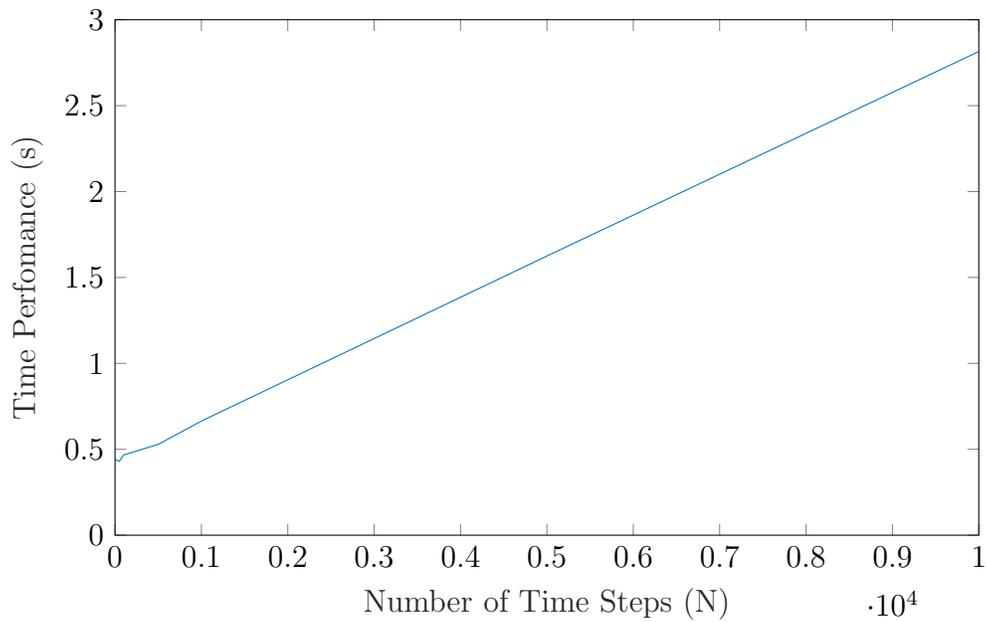
\begin{figure}[!h]
    \centering
    \input{tikz/time.tex}
    \caption{Time Perforance as Function of Number of Time Steps}
    \label{fig:time}
\end{figure}

To calculate how different time step sizes influence the overall efficiency of the program, the model is ran and timed at a different number of time steps $N$. At each number of time steps, the model is ran and timed 5 times. The average time for these five runs is used a measure for the time performance at that specific number of time steps. The results can be see in \autoref{fig:time}. The relationship between the number of time step appears to be linear. When $N=10,000$, the time it takes to run the model has increased by $550 \%$, compared to $N=5$. 

\subsubsection{Impact of Dimensionality}
Next, the effect of changing the number of different stocks $m$ on the efficiency of the program is analyzed. Similarly to the different time step sizes, for different numbers the model is run and timed 5 times. Except for the number of different stocks, all other parameters are the same to allow for fair comparison.\\

\begin{figure}[!h]
    \centering
    \input{tikz/dime.tex}
    \caption{Time Performance as Function of Dimensionality}
    \label{fig:dimper}
\end{figure}
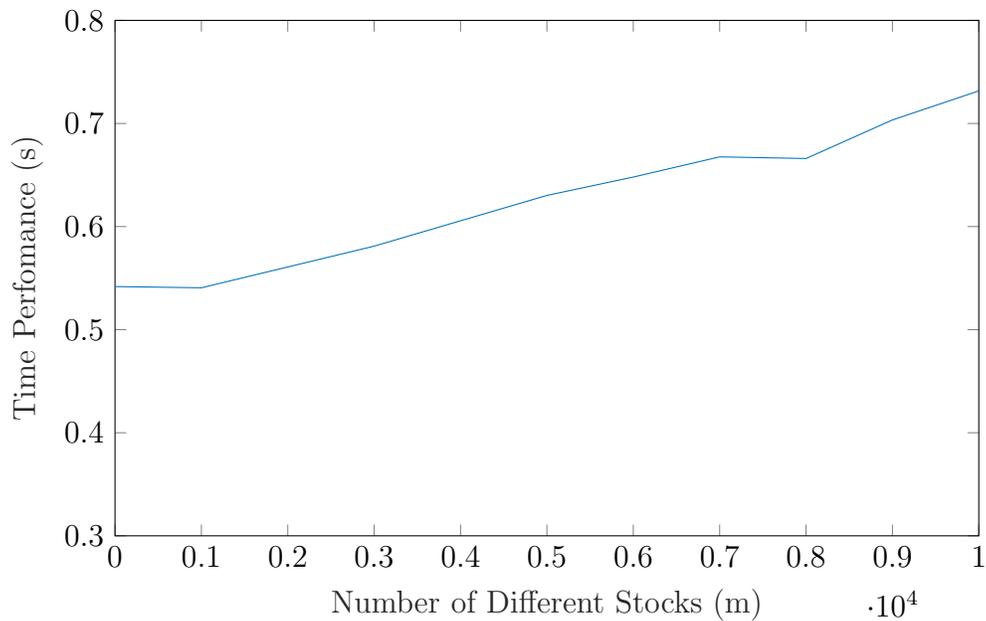

The average of these 5 runs is used as a measure of the model performance for that specific number of different stocks. As can be seen in \autoref{fig:dimper}, the relationship between the number of different stocks and the time performance of the model seems to be linear. At $m=10,000$, the time it takes for the model to run, is approximately $35\%$ higher compared to $m=5$. This means that if the appropriate starting conditions are defined, a set of 10,000 different independently behaving stock prices can be matched at the terminal time, and that it takes only $35 \%$ more effort time wise to calculate. 

%% file: tikz/rendering1.tex
%
%
\definecolor{mycolor1}{rgb}{0.00000,0.44700,0.74100}%
\definecolor{mycolor2}{rgb}{0.85000,0.32500,0.09800}%
\begin{tikzpicture}

\begin{axis}[%
width=4.521in,
height=3.0in,
at={(0.758in,0.481in)},
scale only axis,
unbounded coords=jump,
xmin=0,
xmax=1,
xlabel style={font=\color{white!15!black}},
xlabel={Time (years)},
ymin=40,
ymax=240,
ylabel style={font=\color{white!15!black}},
ylabel={Value (\$)},
axis background/.style={fill=white},
legend style={legend cell align=left, align=left, draw=white!15!black}
]
\addplot [color=mycolor1]
  table[row sep=crcr]{%
0	150\\
0.00273972602739726	152.110708824218\\
0.00547945205479452	159.411256878477\\
0.00821917808219178	149.987385567095\\
0.010958904109589	153.371716993429\\
0.0136986301369863	154.651215574985\\
0.0164383561643836	149.358468425613\\
0.0191780821917808	147.663605004934\\
0.0219178082191781	148.987688619964\\
0.0246575342465753	162.940523940479\\
0.0273972602739726	174.750368334055\\
0.0301369863013699	168.576760084345\\
0.0328767123287671	181.966403722786\\
0.0356164383561644	185.420980624119\\
0.0383561643835616	185.114994944026\\
0.0410958904109589	188.577696195142\\
0.0438356164383562	187.566126413828\\
0.0465753424657534	186.956722927668\\
0.0493150684931507	194.245636143099\\
0.0520547945205479	201.408661139741\\
0.0547945205479452	208.878831968\\
0.0575342465753425	212.549644160629\\
0.0602739726027397	205.832782707839\\
0.063013698630137	209.696468516761\\
0.0657534246575343	218.643211855945\\
0.0684931506849315	221.440737485025\\
0.0712328767123288	227.437165250025\\
0.073972602739726	231.763805562175\\
0.0767123287671233	229.923273705648\\
0.0794520547945206	231.691606200706\\
0.0821917808219178	226.917801608819\\
0.0849315068493151	232.193725889779\\
0.0876712328767123	225.223224555895\\
0.0904109589041096	218.922916017856\\
0.0931506849315069	214.2849138318\\
0.0958904109589041	197.773095883003\\
0.0986301369863014	205.218092559214\\
0.101369863013699	206.964626867737\\
0.104109589041096	202.875548687375\\
0.106849315068493	210.151142995923\\
0.10958904109589	200.737962239468\\
0.112328767123288	200.200826173008\\
0.115068493150685	198.93576558472\\
0.117808219178082	200.597679382127\\
0.120547945205479	202.240149850074\\
0.123287671232877	197.662455799315\\
0.126027397260274	197.50699855392\\
0.128767123287671	196.654739827593\\
0.131506849315069	199.885354500252\\
0.134246575342466	205.604488304378\\
0.136986301369863	211.573394061588\\
0.13972602739726	206.791237154859\\
0.142465753424658	207.209902546688\\
0.145205479452055	200.625829593813\\
0.147945205479452	194.779262629988\\
0.150684931506849	194.744347452005\\
0.153424657534247	202.555692868334\\
0.156164383561644	198.475597831999\\
0.158904109589041	200.404667064687\\
0.161643835616438	199.221515258633\\
0.164383561643836	205.047259713348\\
0.167123287671233	199.202979082542\\
0.16986301369863	199.372713522726\\
0.172602739726027	202.255701383724\\
0.175342465753425	208.081532426088\\
0.178082191780822	216.490912922765\\
0.180821917808219	216.977783783553\\
0.183561643835616	208.507686574873\\
0.186301369863014	204.457022975053\\
0.189041095890411	198.776621362961\\
0.191780821917808	211.004221511279\\
0.194520547945205	207.604724569485\\
0.197260273972603	211.66922320078\\
0.2	210.603294547189\\
0.202739726027397	215.501084286353\\
0.205479452054795	211.187394905842\\
0.208219178082192	203.437016826626\\
0.210958904109589	195.864017054185\\
0.213698630136986	198.366496890534\\
0.216438356164384	197.445655805534\\
0.219178082191781	196.432571172558\\
0.221917808219178	203.729067734865\\
0.224657534246575	205.283747035776\\
0.227397260273973	206.346491115269\\
0.23013698630137	214.920597269985\\
0.232876712328767	210.395694192592\\
0.235616438356164	214.231522074294\\
0.238356164383562	218.913606865393\\
0.241095890410959	217.517306542424\\
0.243835616438356	218.745049074898\\
0.246575342465753	212.070802906745\\
0.249315068493151	205.699496721285\\
0.252054794520548	206.264079463859\\
0.254794520547945	210.162935058549\\
0.257534246575342	224.383693843546\\
0.26027397260274	220.467451964673\\
0.263013698630137	221.54833266745\\
0.265753424657534	221.070014738701\\
0.268493150684932	209.886188385336\\
0.271232876712329	207.474959328959\\
0.273972602739726	197.730090382071\\
0.276712328767123	202.078893385745\\
0.279452054794521	197.382405752692\\
0.282191780821918	197.899458669311\\
0.284931506849315	195.079198742998\\
0.287671232876712	196.628810796391\\
0.29041095890411	193.539523600763\\
0.293150684931507	196.021276754106\\
0.295890410958904	199.814291854147\\
0.298630136986301	208.766401913794\\
0.301369863013699	207.705775664367\\
0.304109589041096	196.081871882162\\
0.306849315068493	191.773356720038\\
0.30958904109589	198.57198184508\\
0.312328767123288	193.000134177064\\
0.315068493150685	197.853954805429\\
0.317808219178082	198.496293930934\\
0.320547945205479	205.9597682972\\
0.323287671232877	195.390098666327\\
0.326027397260274	194.379150482146\\
0.328767123287671	188.234677595596\\
0.331506849315069	202.560475101534\\
0.334246575342466	206.93516670017\\
0.336986301369863	214.403321549643\\
0.33972602739726	208.465669608497\\
0.342465753424658	205.908998630458\\
0.345205479452055	204.440692375758\\
0.347945205479452	210.317766563513\\
0.350684931506849	208.788285109636\\
0.353424657534247	212.621675417098\\
0.356164383561644	201.204213184894\\
0.358904109589041	199.340925816279\\
0.361643835616438	195.044282766465\\
0.364383561643836	186.994128676127\\
0.367123287671233	189.480085509667\\
0.36986301369863	190.878422058181\\
0.372602739726027	191.045671366388\\
0.375342465753425	184.377423372502\\
0.378082191780822	189.817999479409\\
0.380821917808219	191.557608181612\\
0.383561643835616	190.058302923877\\
0.386301369863014	190.172157968972\\
0.389041095890411	188.868198084016\\
0.391780821917808	180.21706172091\\
0.394520547945205	178.869790036509\\
0.397260273972603	174.977966896824\\
0.4	170.493806493002\\
0.402739726027397	165.333911257669\\
0.405479452054795	163.025216998938\\
0.408219178082192	154.480841604389\\
0.410958904109589	158.379171409592\\
0.413698630136986	160.534806154515\\
0.416438356164384	160.450661381011\\
0.419178082191781	160.304651132809\\
0.421917808219178	156.956063992657\\
0.424657534246575	161.140545982488\\
0.427397260273973	160.578735788195\\
0.43013698630137	157.575895916733\\
0.432876712328767	163.148942820371\\
0.435616438356164	162.189214695223\\
0.438356164383562	159.688971563435\\
0.441095890410959	158.461301430458\\
0.443835616438356	154.944848021799\\
0.446575342465753	150.402624700796\\
0.449315068493151	160.345506593826\\
0.452054794520548	167.292692485458\\
0.454794520547945	168.639158255205\\
0.457534246575342	163.090880850901\\
0.46027397260274	159.396817997163\\
0.463013698630137	158.660387335383\\
0.465753424657534	161.946609664362\\
0.468493150684932	156.301126728605\\
0.471232876712329	146.770601397495\\
0.473972602739726	141.204383659047\\
0.476712328767123	142.436868813474\\
0.479452054794521	143.895734562853\\
0.482191780821918	145.596724697196\\
0.484931506849315	145.100282786417\\
0.487671232876712	145.797832568634\\
0.49041095890411	143.980973754967\\
0.493150684931507	147.229204199886\\
0.495890410958904	141.982371569499\\
0.498630136986301	143.67319246121\\
0.501369863013699	140.481960917129\\
0.504109589041096	139.250721976321\\
0.506849315068493	141.265262361068\\
0.50958904109589	145.106864556561\\
0.512328767123288	140.862497303226\\
0.515068493150685	145.509966116489\\
0.517808219178082	148.023903330852\\
0.520547945205479	147.760994714197\\
0.523287671232877	147.006058339064\\
0.526027397260274	146.168856121823\\
0.528767123287671	145.009344043737\\
0.531506849315069	145.096803755154\\
0.534246575342466	145.291571542294\\
0.536986301369863	148.432637480622\\
0.53972602739726	154.364422228827\\
0.542465753424658	156.250710590404\\
0.545205479452055	155.393136548186\\
0.547945205479452	157.935675855621\\
0.550684931506849	158.693021112087\\
0.553424657534247	154.416204781627\\
0.556164383561644	158.252253943278\\
0.558904109589041	159.523997178957\\
0.561643835616438	160.088343616011\\
0.564383561643836	162.247072959801\\
0.567123287671233	163.357056484606\\
0.56986301369863	159.331969852442\\
0.572602739726027	158.655036654138\\
0.575342465753425	158.048589758043\\
0.578082191780822	155.848020119141\\
0.580821917808219	162.708865001479\\
0.583561643835616	158.979759321497\\
0.586301369863014	156.966754657216\\
0.589041095890411	154.04183447175\\
0.591780821917808	149.308045527258\\
0.594520547945205	148.556856128176\\
0.597260273972603	147.491295571113\\
0.6	153.397411622897\\
0.602739726027397	152.397678449166\\
0.605479452054795	148.153138244448\\
0.608219178082192	154.370295122585\\
0.610958904109589	159.358467117016\\
0.613698630136986	158.400786650766\\
0.616438356164384	152.156944281754\\
0.619178082191781	150.386378856847\\
0.621917808219178	149.772627466997\\
0.624657534246575	150.854740154982\\
0.627397260273973	149.823653286615\\
0.63013698630137	151.562339081452\\
0.632876712328767	153.116813995167\\
0.635616438356164	148.10503073343\\
0.638356164383562	144.430652856704\\
0.641095890410959	141.629328929002\\
0.643835616438356	139.747047854666\\
0.646575342465753	138.574591859763\\
0.649315068493151	138.619812868497\\
0.652054794520548	127.630423028096\\
0.654794520547945	126.103884063304\\
0.657534246575342	130.204323471481\\
0.66027397260274	126.569428705286\\
0.663013698630137	129.662378353413\\
0.665753424657534	130.851164482762\\
0.668493150684931	130.751833331566\\
0.671232876712329	131.37617221911\\
0.673972602739726	125.99508462037\\
0.676712328767123	125.716320582962\\
0.679452054794521	130.993539026769\\
0.682191780821918	131.330700532776\\
0.684931506849315	131.472905015178\\
0.687671232876712	128.946773595447\\
0.69041095890411	128.842786768791\\
0.693150684931507	129.626253888332\\
0.695890410958904	131.072762639393\\
0.698630136986301	129.793905263688\\
0.701369863013699	128.990701732258\\
0.704109589041096	135.822360305323\\
0.706849315068493	127.794739332623\\
0.70958904109589	135.251212235968\\
0.712328767123288	136.446081741873\\
0.715068493150685	140.01725609239\\
0.717808219178082	133.91906005345\\
0.720547945205479	131.851093940242\\
0.723287671232877	130.891577696929\\
0.726027397260274	132.339627982988\\
0.728767123287671	126.554910648107\\
0.731506849315068	128.117007582054\\
0.734246575342466	124.050360838901\\
0.736986301369863	124.265249897263\\
0.73972602739726	126.386822295869\\
0.742465753424657	127.468636689798\\
0.745205479452055	131.080309304017\\
0.747945205479452	134.531685489292\\
0.750684931506849	132.239933647965\\
0.753424657534247	133.129573452541\\
0.756164383561644	129.839209906986\\
0.758904109589041	125.347709821804\\
0.761643835616438	128.381600832489\\
0.764383561643836	128.3817683204\\
0.767123287671233	128.197246015464\\
0.76986301369863	131.254146571128\\
0.772602739726027	133.296587956823\\
0.775342465753425	134.518275659534\\
0.778082191780822	138.919790328943\\
0.780821917808219	142.300223147477\\
0.783561643835616	143.19314156904\\
0.786301369863014	140.605988625447\\
0.789041095890411	138.208384004288\\
0.791780821917808	142.520310317645\\
0.794520547945206	136.50830064155\\
0.797260273972603	136.420908281193\\
0.8	129.462942427359\\
0.802739726027397	132.920593858881\\
0.805479452054795	135.91823854557\\
0.808219178082192	135.922372242558\\
0.810958904109589	135.670386490789\\
0.813698630136986	126.842446839528\\
0.816438356164384	128.771717264829\\
0.819178082191781	121.38297070507\\
0.821917808219178	114.015225905165\\
0.824657534246575	114.253741289454\\
0.827397260273973	111.417634168947\\
0.83013698630137	112.617513150792\\
0.832876712328767	114.61279354815\\
0.835616438356164	117.185608548334\\
0.838356164383562	115.06590123517\\
0.841095890410959	116.419162523009\\
0.843835616438356	116.725775106511\\
0.846575342465753	119.249297646499\\
0.849315068493151	120.922588538489\\
0.852054794520548	123.764125966909\\
0.854794520547945	123.336771927821\\
0.857534246575343	122.861624638712\\
0.86027397260274	126.102055198659\\
0.863013698630137	119.093479057309\\
0.865753424657534	117.520774663309\\
0.868493150684931	113.612859099644\\
0.871232876712329	112.475287408582\\
0.873972602739726	114.384747991655\\
0.876712328767123	116.856632956188\\
0.879452054794521	113.752653284499\\
0.882191780821918	112.350257898736\\
0.884931506849315	112.753157704101\\
0.887671232876712	111.891902445228\\
0.89041095890411	112.775731279555\\
0.893150684931507	113.956117772148\\
0.895890410958904	111.182630262745\\
0.898630136986301	110.668092663125\\
0.901369863013699	104.492872197991\\
0.904109589041096	107.625096822026\\
0.906849315068493	105.853151019138\\
0.90958904109589	102.518121130893\\
0.912328767123288	101.836781995322\\
0.915068493150685	98.0291660941469\\
0.917808219178082	97.975655030527\\
0.920547945205479	96.5380314393897\\
0.923287671232877	102.040223658518\\
0.926027397260274	105.080512218185\\
0.928767123287671	98.2138678552678\\
0.931506849315068	99.348243301851\\
0.934246575342466	95.7129976360603\\
0.936986301369863	95.0741038881872\\
0.93972602739726	95.483174599869\\
0.942465753424658	97.3516938114554\\
0.945205479452055	96.656021988952\\
0.947945205479452	100.643436685476\\
0.950684931506849	99.376667888092\\
0.953424657534247	100.228463464136\\
0.956164383561644	101.972126721707\\
0.958904109589041	102.199472290132\\
0.961643835616438	104.555741230055\\
0.964383561643836	105.440165115244\\
0.967123287671233	103.27631745305\\
0.96986301369863	98.3966374436726\\
0.972602739726027	103.182807380778\\
0.975342465753425	101.55032375633\\
0.978082191780822	101.82502221228\\
0.980821917808219	103.325796047267\\
0.983561643835616	103.632980464876\\
0.986301369863014	101.179182883472\\
0.989041095890411	99.9406839114291\\
0.991780821917808	99.6140258568716\\
0.994520547945206	103.469697573025\\
0.997260273972603	101.138672601218\\
1	103.215627289625\\
};
\addlegendentry{Stock Price}

\addplot [color=mycolor2]
  table[row sep=crcr]{%
0	50\\
0.00273972602739726	50.2781204897294\\
0.00547945205479452	50.5763170526602\\
0.00821917808219178	50.8485391031121\\
0.010958904109589	51.1301226312507\\
0.0136986301369863	51.4152587302675\\
0.0164383561643836	51.6856866957884\\
0.0191780821917808	51.9513953523436\\
0.0219178082191781	52.2208112568619\\
0.0246575342465753	52.5293496820139\\
0.0273972602739726	52.8710964599923\\
0.0301369863013699	53.195442967884\\
0.0328767123287671	53.5576536585546\\
0.0356164383561644	53.9296656251654\\
0.0383561643835616	54.3008133126139\\
0.0410958904109589	54.6818417483893\\
0.0438356164383562	55.0599830129573\\
0.0465753424657534	55.4363823094132\\
0.0493150684931507	55.8337545090702\\
0.0520547945205479	56.2517973404972\\
0.0547945205479452	56.6914596678456\\
0.0575342465753425	57.1417798756453\\
0.0602739726027397	57.5725585898821\\
0.063013698630137	58.014620791795\\
0.0657534246575343	58.4828797782495\\
0.0684931506849315	58.9593586926336\\
0.0712328767123288	59.4535017946828\\
0.073972602739726	59.960429927628\\
0.0767123287671233	60.4619123882725\\
0.0794520547945206	60.9686553171276\\
0.0821917808219178	61.4611787654261\\
0.0849315068493151	61.9694785210572\\
0.0876712328767123	62.4568874258604\\
0.0904109589041096	62.9253576498857\\
0.0931506849315069	63.3798456033145\\
0.0958904109589041	63.784386200087\\
0.0986301369863014	64.2115262218962\\
0.101369863013699	64.6439894254887\\
0.104109589041096	65.0639747415728\\
0.106849315068493	65.5062495180478\\
0.10958904109589	65.9196132746634\\
0.112328767123288	66.3313289159012\\
0.115068493150685	66.7391411964272\\
0.117808219178082	67.152113979873\\
0.120547945205479	67.57020294586\\
0.123287671232877	67.9740155811597\\
0.126027397260274	68.377348969101\\
0.128767123287671	68.7780136853336\\
0.131506849315069	69.1888619686128\\
0.134246575342466	69.617789679612\\
0.136986301369863	70.0656462134079\\
0.13972602739726	70.498302539519\\
0.142465753424658	70.932302156036\\
0.145205479452055	71.3452366754972\\
0.147945205479452	71.7394062007767\\
0.150684931506849	72.1334710083841\\
0.153424657534247	72.5527879161432\\
0.156164383561644	72.958883867078\\
0.158904109589041	73.3712627237579\\
0.161643835616438	73.7797882469661\\
0.164383561643836	74.2073968447615\\
0.167123287671233	74.6158145138371\\
0.16986301369863	75.0247996861258\\
0.172602739726027	75.4433267523365\\
0.175342465753425	75.8811916934256\\
0.178082191780822	76.3470599109628\\
0.180821917808219	76.8145626263776\\
0.183561643835616	77.2536865879966\\
0.186301369863014	77.6791974726514\\
0.189041095890411	78.0855501635435\\
0.191780821917808	78.5333094845738\\
0.194520547945205	78.9695285682744\\
0.197260273972603	79.4196113205691\\
0.2	79.8660566191155\\
0.202739726027397	80.3293213206118\\
0.205479452054795	80.7777375419106\\
0.208219178082192	81.199376079819\\
0.210958904109589	81.5947588720979\\
0.213698630136986	81.9988598637053\\
0.216438356164384	82.399753676618\\
0.219178082191781	82.7971057418991\\
0.221917808219178	83.2201297895299\\
0.224657534246575	83.6486502844962\\
0.227397260273973	84.0809442608439\\
0.23013698630137	84.5437268177009\\
0.232876712328767	84.9903756397095\\
0.235616438356164	85.4507673893499\\
0.238356164383562	85.927992511742\\
0.241095890410959	86.4001914321614\\
0.243835616438356	86.8768433960762\\
0.246575342465753	87.3292598275091\\
0.249315068493151	87.7584565392467\\
0.252054794520548	88.1897287103631\\
0.254794520547945	88.635330062182\\
0.257534246575342	89.1333639901654\\
0.26027397260274	89.616917220583\\
0.263013698630137	90.1044943677999\\
0.265753424657534	90.5902982069058\\
0.268493150684932	91.0342648368325\\
0.271232876712329	91.4691851027249\\
0.273972602739726	91.8673765681288\\
0.276712328767123	92.2820354475647\\
0.279452054794521	92.6788633014271\\
0.282191780821918	93.0776728669888\\
0.284931506849315	93.4656965997476\\
0.287671232876712	93.859685186875\\
0.29041095890411	94.2417666537248\\
0.293150684931507	94.6334691394784\\
0.295890410958904	95.0399279686765\\
0.298630136986301	95.4813364847228\\
0.301369863013699	95.9185995046834\\
0.304109589041096	96.3101476055225\\
0.306849315068493	96.6846904399365\\
0.30958904109589	97.0862004179944\\
0.312328767123288	97.4655401765874\\
0.315068493150685	97.8642902867679\\
0.317808219178082	98.2656286183405\\
0.320547945205479	98.6970487925388\\
0.323287671232877	99.0857203892285\\
0.326027397260274	99.4702962772824\\
0.328767123287671	99.8298222965872\\
0.331506849315069	100.24802775347\\
0.334246575342466	100.684233669243\\
0.336986301369863	101.151288214825\\
0.33972602739726	101.593733800438\\
0.342465753424658	102.025545117898\\
0.345205479452055	102.451228221452\\
0.347945205479452	102.901598919384\\
0.350684931506849	103.345531637028\\
0.353424657534247	103.805707754036\\
0.356164383561644	104.21734248713\\
0.358904109589041	104.6210309517\\
0.361643835616438	105.006302161143\\
0.364383561643836	105.356908003713\\
0.367123287671233	105.718280149199\\
0.36986301369863	106.085739652948\\
0.372602739726027	106.453940520018\\
0.375342465753425	106.7929239546\\
0.378082191780822	107.155871867643\\
0.380821917808219	107.52652419738\\
0.383561643835616	107.890528454619\\
0.386301369863014	108.255052362799\\
0.389041095890411	108.613743932343\\
0.391780821917808	108.933499933065\\
0.394520547945205	109.247174813895\\
0.397260273972603	109.543181000765\\
0.4	109.818734362874\\
0.402739726027397	110.070642826318\\
0.405479452054795	110.311929567336\\
0.408219178082192	110.513691169881\\
0.410958904109589	110.733587351286\\
0.413698630136986	110.963563565242\\
0.416438356164384	111.193156920019\\
0.419178082191781	111.422073896864\\
0.421917808219178	111.635140342742\\
0.424657534246575	111.868135685251\\
0.427397260273973	112.098456263169\\
0.43013698630137	112.314358729647\\
0.432876712328767	112.557184343007\\
0.435616438356164	112.795365276929\\
0.438356164383562	113.021367315835\\
0.441095890410959	113.24136611207\\
0.443835616438356	113.444061878868\\
0.446575342465753	113.624292785452\\
0.449315068493151	113.853983317256\\
0.452054794520548	114.118408480844\\
0.454794520547945	114.389609537606\\
0.457534246575342	114.632811928209\\
0.46027397260274	114.857282062908\\
0.463013698630137	115.078008677341\\
0.465753424657534	115.315594151658\\
0.468493150684932	115.524101821028\\
0.471232876712329	115.683257910677\\
0.473972602739726	115.813445504\\
0.476712328767123	115.950096298847\\
0.479452054794521	116.094435399295\\
0.482191780821918	116.247784164542\\
0.484931506849315	116.398505659125\\
0.487671232876712	116.552968824683\\
0.49041095890411	116.69767940525\\
0.493150684931507	116.859955550973\\
0.495890410958904	116.993736946573\\
0.498630136986301	117.136767910753\\
0.501369863013699	117.262281993154\\
0.504109589041096	117.381008102612\\
0.506849315068493	117.510935938787\\
0.50958904109589	117.66233261145\\
0.512328767123288	117.789903010516\\
0.515068493150685	117.943736843111\\
0.517808219178082	118.111863884818\\
0.520547945205479	118.278501692816\\
0.523287671232877	118.440814510077\\
0.526027397260274	118.598301827942\\
0.528767123287671	118.74906199146\\
0.531506849315069	118.900346349376\\
0.534246575342466	119.052789032112\\
0.536986301369863	119.223827320929\\
0.53972602739726	119.430180397219\\
0.542465753424658	119.647839561764\\
0.545205479452055	119.860346415837\\
0.547945205479452	120.088273064031\\
0.550684931506849	120.32082993219\\
0.553424657534247	120.527165599219\\
0.556164383561644	120.757190200848\\
0.558904109589041	120.995125790019\\
0.561643835616438	121.236601155678\\
0.564383561643836	121.491664820657\\
0.567123287671233	121.753765980222\\
0.56986301369863	121.990245759612\\
0.572602739726027	122.222399884845\\
0.575342465753425	122.450655067085\\
0.578082191780822	122.664635477969\\
0.580821917808219	122.923467105094\\
0.583561643835616	123.157780573096\\
0.586301369863014	123.378777288694\\
0.589041095890411	123.580289364355\\
0.591780821917808	123.750046662129\\
0.594520547945205	123.914742065662\\
0.597260273972603	124.072202589993\\
0.6	124.270127141615\\
0.602739726027397	124.461170758246\\
0.605479452054795	124.622753173181\\
0.608219178082192	124.827823958767\\
0.610958904109589	125.068034998246\\
0.613698630136986	125.3014676954\\
0.616438356164384	125.49031614556\\
0.619178082191781	125.666440524391\\
0.621917808219178	125.838131075903\\
0.624657534246575	126.017733788077\\
0.627397260273973	126.189768633571\\
0.63013698630137	126.374696278757\\
0.632876712328767	126.571238204248\\
0.635616438356164	126.730110848039\\
0.638356164383562	126.861160579019\\
0.641095890410959	126.9708394501\\
0.643835616438356	127.066052607595\\
0.646575342465753	127.152190429609\\
0.649315068493151	127.23869529624\\
0.652054794520548	127.238679979383\\
0.654794520547945	127.226562508523\\
0.657534246575342	127.247263856354\\
0.66027397260274	127.238663772418\\
0.663013698630137	127.255224787741\\
0.665753424657534	127.281544297422\\
0.668493150684931	127.307056592698\\
0.671232876712329	127.337785833311\\
0.673972602739726	127.323306282425\\
0.676712328767123	127.3064779024\\
0.679452054794521	127.334771420494\\
0.682191780821918	127.365985517695\\
0.684931506849315	127.398450054282\\
0.687671232876712	127.408767215877\\
0.69041095890411	127.418177820527\\
0.693150684931507	127.434598158829\\
0.695890410958904	127.464065273722\\
0.698630136986301	127.481918901562\\
0.701369863013699	127.492416623326\\
0.704109589041096	127.566191729351\\
0.706849315068493	127.564946935638\\
0.70958904109589	127.634068903577\\
0.712328767123288	127.714585858106\\
0.715068493150685	127.829459431105\\
0.717808219178082	127.885132883465\\
0.720547945205479	127.920543115116\\
0.723287671232877	127.946465634909\\
0.726027397260274	127.986884639375\\
0.728767123287671	127.968877094789\\
0.731506849315068	127.966825629692\\
0.734246575342466	127.922857125091\\
0.736986301369863	127.881141239418\\
0.73972602739726	127.861775228905\\
0.742465753424657	127.853933654433\\
0.745205479452055	127.884947806461\\
0.747945205479452	127.953497354146\\
0.750684931506849	127.996872122602\\
0.753424657534247	128.050147420323\\
0.756164383561644	128.066459301803\\
0.758904109589041	128.031735591022\\
0.761643835616438	128.031904740087\\
0.764383561643836	128.032089675757\\
0.767123287671233	128.030117168285\\
0.76986301369863	128.064557512042\\
0.772602739726027	128.123624462471\\
0.775342465753425	128.197606985892\\
0.778082191780822	128.325954497022\\
0.780821917808219	128.49658031034\\
0.783561643835616	128.678525222912\\
0.786301369863014	128.827308146269\\
0.789041095890411	128.944960359822\\
0.791780821917808	129.119375134231\\
0.794520547945206	129.213625502419\\
0.797260273972603	129.306708587383\\
0.8	129.304468859228\\
0.802739726027397	129.350277348876\\
0.805479452054795	129.438330215237\\
0.808219178082192	129.526456139323\\
0.810958904109589	129.610943208684\\
0.813698630136986	129.565589774157\\
0.816438356164384	129.549052564273\\
0.819178082191781	129.420550627226\\
0.821917808219178	129.17868411676\\
0.824657534246575	128.940559322198\\
0.827397260273973	128.657419308126\\
0.83013698630137	128.393651110268\\
0.832876712328767	128.162614867274\\
0.835616438356164	127.974484045348\\
0.838356164383562	127.750430253272\\
0.841095890410959	127.549728618359\\
0.843835616438356	127.354421392003\\
0.846575342465753	127.204203104731\\
0.849315068493151	127.084430382941\\
0.852054794520548	127.017306998812\\
0.854794520547945	126.942131816719\\
0.857534246575343	126.857830348302\\
0.86027397260274	126.837098360364\\
0.863013698630137	126.676168782505\\
0.865753424657534	126.483147723577\\
0.868493150684931	126.208702037613\\
0.871232876712329	125.910059345056\\
0.873972602739726	125.652952341093\\
0.876712328767123	125.450805753612\\
0.879452054794521	125.178106997711\\
0.882191780821918	124.872798288403\\
0.884931506849315	124.577098778233\\
0.887671232876712	124.260400236531\\
0.89041095890411	123.965817475193\\
0.893150684931507	123.701523556018\\
0.895890410958904	123.364234393753\\
0.898630136986301	123.01304791439\\
0.901369863013699	122.4902884315\\
0.904109589041096	122.057061771816\\
0.906849315068493	121.57171620078\\
0.90958904109589	120.985291010479\\
0.912328767123288	120.377580440913\\
0.915068493150685	119.647018302461\\
0.917808219178082	118.914684836683\\
0.920547945205479	118.132775447416\\
0.923287671232877	117.547447531206\\
0.926027397260274	117.074771147518\\
0.928767123287671	116.337922086922\\
0.931506849315068	115.646475028121\\
0.934246575342466	114.803523666023\\
0.936986301369863	113.932797854561\\
0.93972602739726	113.080684433071\\
0.942465753424658	112.317588889992\\
0.945205479452055	111.519710709098\\
0.947945205479452	110.931776155979\\
0.950684931506849	110.273454818701\\
0.953424657534247	109.665267237605\\
0.956164383561644	109.166105795427\\
0.958904109589041	108.682117446287\\
0.961643835616438	108.366500516406\\
0.964383561643836	108.118950069548\\
0.967123287671233	107.691031845196\\
0.96986301369863	106.819373087168\\
0.972602739726027	106.426499385631\\
0.975342465753425	105.852190738975\\
0.978082191780822	105.312242091878\\
0.980821917808219	104.986771398609\\
0.983561643835616	104.712526257643\\
0.986301369863014	103.94737176752\\
0.989041095890411	102.872502047933\\
0.991780821917808	101.688721643057\\
0.994520547945206	102.433421861855\\
0.997260273972603	100.846341732293\\
1	inf\\
};
\addlegendentry{Accumulated Cash}

\end{axis}
\end{tikzpicture}%

%% file: tikz/22.tex
%
%
\definecolor{mycolor1}{rgb}{0.00000,0.44700,0.74100}%
\begin{tikzpicture}

\begin{axis}[%
width=2.6in,
height=2.6in,
at={(0.758in,0.481in)},
scale only axis,
xmin=0,
xmax=1,
xlabel style={font=\color{white!15!black}},
xlabel={Time (years)},
ymin=0,
ymax=700,
ylabel style={font=\color{white!15!black}},
ylabel={Value (\$)},
axis background/.style={fill=white},
legend style={legend cell align=left, align=left, draw=white!15!black}
]
\addplot [color=mycolor1]
  table[row sep=crcr]{%
0	200\\
0.00273972602739726	202.493001843572\\
0.00547945205479452	204.635322955394\\
0.00821917808219178	207.207886569587\\
0.010958904109589	205.103835820726\\
0.0136986301369863	201.797122968383\\
0.0164383561643836	209.121442858542\\
0.0191780821917808	206.318699086181\\
0.0219178082191781	215.986532438912\\
0.0246575342465753	214.836103473744\\
0.0273972602739726	219.988277015401\\
0.0301369863013699	217.001320987722\\
0.0328767123287671	220.908152613836\\
0.0356164383561644	223.669560191187\\
0.0383561643835616	221.5724105691\\
0.0410958904109589	224.444552161437\\
0.0438356164383562	235.082639227542\\
0.0465753424657534	226.740791008914\\
0.0493150684931507	227.396852406621\\
0.0520547945205479	239.474037221213\\
0.0547945205479452	236.030790066713\\
0.0575342465753425	238.460061342682\\
0.0602739726027397	235.724593256496\\
0.063013698630137	219.22714221172\\
0.0657534246575343	221.681155869442\\
0.0684931506849315	213.019923356647\\
0.0712328767123288	206.538447812565\\
0.073972602739726	207.348172892676\\
0.0767123287671233	212.417614000904\\
0.0794520547945206	219.456260501331\\
0.0821917808219178	217.813992788553\\
0.0849315068493151	213.660082137139\\
0.0876712328767123	206.262213842595\\
0.0904109589041096	199.717076274493\\
0.0931506849315069	204.211881379171\\
0.0958904109589041	213.11838928365\\
0.0986301369863014	211.454830410596\\
0.101369863013699	211.768991656601\\
0.104109589041096	213.867084920427\\
0.106849315068493	227.125138885677\\
0.10958904109589	234.082190436946\\
0.112328767123288	232.758963796617\\
0.115068493150685	230.473926972577\\
0.117808219178082	227.212676050316\\
0.120547945205479	225.973467423855\\
0.123287671232877	207.831785952124\\
0.126027397260274	199.855490004262\\
0.128767123287671	199.988125956686\\
0.131506849315069	207.087540979115\\
0.134246575342466	198.879656605717\\
0.136986301369863	211.738597409067\\
0.13972602739726	214.83150819507\\
0.142465753424658	213.12288087665\\
0.145205479452055	214.137164305128\\
0.147945205479452	204.523847546041\\
0.150684931506849	206.261788084164\\
0.153424657534247	208.040494806193\\
0.156164383561644	203.823180837498\\
0.158904109589041	196.989679807654\\
0.161643835616438	198.280108115317\\
0.164383561643836	192.73710954347\\
0.167123287671233	189.630322668728\\
0.16986301369863	192.209609486993\\
0.172602739726027	190.85006031321\\
0.175342465753425	189.400578501042\\
0.178082191780822	189.272484328591\\
0.180821917808219	185.050834266601\\
0.183561643835616	185.536942982283\\
0.186301369863014	185.744170569343\\
0.189041095890411	186.2702020523\\
0.191780821917808	194.060569916521\\
0.194520547945205	193.618694725857\\
0.197260273972603	193.031630818049\\
0.2	191.215878185011\\
0.202739726027397	187.715722606013\\
0.205479452054795	196.617658072914\\
0.208219178082192	202.583767961694\\
0.210958904109589	212.806242816488\\
0.213698630136986	217.615019966286\\
0.216438356164384	217.575259044785\\
0.219178082191781	223.559173385466\\
0.221917808219178	216.894179952751\\
0.224657534246575	209.542964433679\\
0.227397260273973	207.138358764598\\
0.23013698630137	203.653716762112\\
0.232876712328767	197.50640673122\\
0.235616438356164	201.28587947952\\
0.238356164383562	203.215494466789\\
0.241095890410959	194.329973920162\\
0.243835616438356	204.79633875329\\
0.246575342465753	197.14295555934\\
0.249315068493151	194.434170636017\\
0.252054794520548	192.018206603679\\
0.254794520547945	194.11769487812\\
0.257534246575342	183.741553058661\\
0.26027397260274	183.347348909224\\
0.263013698630137	178.459208820866\\
0.265753424657534	178.065821949258\\
0.268493150684932	172.695843875225\\
0.271232876712329	170.665272415755\\
0.273972602739726	167.186291368176\\
0.276712328767123	178.901190996657\\
0.279452054794521	174.056645677829\\
0.282191780821918	172.212428745733\\
0.284931506849315	168.663393953966\\
0.287671232876712	165.80612224605\\
0.29041095890411	161.70314984654\\
0.293150684931507	166.264933699242\\
0.295890410958904	169.246691987465\\
0.298630136986301	166.548385543214\\
0.301369863013699	165.650984286113\\
0.304109589041096	167.089980711303\\
0.306849315068493	163.982226452149\\
0.30958904109589	164.246871181683\\
0.312328767123288	159.596670366827\\
0.315068493150685	157.433500053601\\
0.317808219178082	159.487888519346\\
0.320547945205479	160.597839044516\\
0.323287671232877	162.143897346409\\
0.326027397260274	164.708353758846\\
0.328767123287671	164.507117322441\\
0.331506849315069	164.64463218453\\
0.334246575342466	165.19259817416\\
0.336986301369863	166.625695566625\\
0.33972602739726	162.561424474097\\
0.342465753424658	160.848665434748\\
0.345205479452055	152.735111778828\\
0.347945205479452	158.634466140218\\
0.350684931506849	158.289269342087\\
0.353424657534247	163.686747491833\\
0.356164383561644	169.033312299586\\
0.358904109589041	163.141865454378\\
0.361643835616438	164.339462887003\\
0.364383561643836	164.281755653583\\
0.367123287671233	166.744951143793\\
0.36986301369863	157.493594544618\\
0.372602739726027	155.170548238084\\
0.375342465753425	153.033473730159\\
0.378082191780822	150.523907870528\\
0.380821917808219	157.24604370138\\
0.383561643835616	158.362935891304\\
0.386301369863014	164.080422692381\\
0.389041095890411	162.427312401271\\
0.391780821917808	162.075236140355\\
0.394520547945205	160.477583496441\\
0.397260273972603	154.233429573813\\
0.4	150.099153607894\\
0.402739726027397	155.840393119249\\
0.405479452054795	159.145266788353\\
0.408219178082192	165.002298253854\\
0.410958904109589	174.449870151323\\
0.413698630136986	170.973310627909\\
0.416438356164384	171.473108290887\\
0.419178082191781	172.851152871086\\
0.421917808219178	171.767256263743\\
0.424657534246575	170.142008093995\\
0.427397260273973	167.404582659143\\
0.43013698630137	166.200321916985\\
0.432876712328767	165.855092195454\\
0.435616438356164	163.659345834112\\
0.438356164383562	157.83008379665\\
0.441095890410959	163.390048751917\\
0.443835616438356	168.233010921698\\
0.446575342465753	173.688156085252\\
0.449315068493151	177.874127809827\\
0.452054794520548	174.408192793692\\
0.454794520547945	178.516620078369\\
0.457534246575342	176.403553027844\\
0.46027397260274	169.095212522533\\
0.463013698630137	171.260190705152\\
0.465753424657534	171.133739763111\\
0.468493150684932	173.867597439068\\
0.471232876712329	181.876935405116\\
0.473972602739726	191.083802191522\\
0.476712328767123	195.382273995461\\
0.479452054794521	193.911938729364\\
0.482191780821918	184.747398627088\\
0.484931506849315	185.594117069912\\
0.487671232876712	187.13684454332\\
0.49041095890411	188.964465923869\\
0.493150684931507	187.012569785054\\
0.495890410958904	187.587446033263\\
0.498630136986301	189.506020901694\\
0.501369863013699	183.403014856699\\
0.504109589041096	184.839981376495\\
0.506849315068493	182.476290913932\\
0.50958904109589	184.7088115828\\
0.512328767123288	189.367607842556\\
0.515068493150685	194.352213169973\\
0.517808219178082	183.64390862087\\
0.520547945205479	176.874758811598\\
0.523287671232877	180.236609251997\\
0.526027397260274	177.189416649827\\
0.528767123287671	180.978390744329\\
0.531506849315069	179.14152808841\\
0.534246575342466	181.143623684865\\
0.536986301369863	173.199193630452\\
0.53972602739726	174.837612547389\\
0.542465753424658	174.094198464222\\
0.545205479452055	171.988440992622\\
0.547945205479452	170.337730888171\\
0.550684931506849	173.114592161128\\
0.553424657534247	177.19289966407\\
0.556164383561644	181.290071886327\\
0.558904109589041	182.19529598107\\
0.561643835616438	187.17611934888\\
0.564383561643836	188.968500299298\\
0.567123287671233	186.813339589579\\
0.56986301369863	188.542642251822\\
0.572602739726027	189.816050670116\\
0.575342465753425	183.190410078517\\
0.578082191780822	184.848178399825\\
0.580821917808219	187.369675952009\\
0.583561643835616	177.353794553706\\
0.586301369863014	169.814982774055\\
0.589041095890411	171.984735722898\\
0.591780821917808	167.144910100985\\
0.594520547945205	162.091284894624\\
0.597260273972603	166.34751678303\\
0.6	174.089054869343\\
0.602739726027397	173.839826290268\\
0.605479452054795	177.104762586096\\
0.608219178082192	172.389938546787\\
0.610958904109589	170.835194729568\\
0.613698630136986	174.961922145376\\
0.616438356164384	169.759523741583\\
0.619178082191781	167.574192820177\\
0.621917808219178	167.307284672299\\
0.624657534246575	169.715416243861\\
0.627397260273973	177.927936373804\\
0.63013698630137	176.634536560569\\
0.632876712328767	183.552442220635\\
0.635616438356164	188.538071513292\\
0.638356164383562	196.967244366106\\
0.641095890410959	198.572875084315\\
0.643835616438356	201.319349318463\\
0.646575342465753	203.093994221465\\
0.649315068493151	194.641241834401\\
0.652054794520548	197.609881220585\\
0.654794520547945	204.013676157121\\
0.657534246575342	209.616712326203\\
0.66027397260274	204.196795151943\\
0.663013698630137	206.003787238291\\
0.665753424657534	210.592350689905\\
0.668493150684931	202.295345861284\\
0.671232876712329	206.387151043737\\
0.673972602739726	204.683042980673\\
0.676712328767123	205.274156698249\\
0.679452054794521	204.926133977891\\
0.682191780821918	202.013572414981\\
0.684931506849315	203.436409303597\\
0.687671232876712	200.785636826377\\
0.69041095890411	196.278572629213\\
0.693150684931507	193.043214601936\\
0.695890410958904	189.664057825747\\
0.698630136986301	185.5138470696\\
0.701369863013699	179.1419097232\\
0.704109589041096	190.525358590829\\
0.706849315068493	190.535038894477\\
0.70958904109589	185.040048624418\\
0.712328767123288	190.719697917541\\
0.715068493150685	190.052630288181\\
0.717808219178082	182.887144601784\\
0.720547945205479	181.19619903201\\
0.723287671232877	176.962316471161\\
0.726027397260274	175.165921689772\\
0.728767123287671	176.37644693623\\
0.731506849315068	177.688383640046\\
0.734246575342466	176.529208649322\\
0.736986301369863	173.579341469479\\
0.73972602739726	170.929576080426\\
0.742465753424657	169.223162209182\\
0.745205479452055	166.242620548662\\
0.747945205479452	169.015135243169\\
0.750684931506849	163.579214575478\\
0.753424657534247	170.124100305241\\
0.756164383561644	171.747004142931\\
0.758904109589041	163.670096321311\\
0.761643835616438	158.966985210549\\
0.764383561643836	165.267581892561\\
0.767123287671233	167.399541651275\\
0.76986301369863	160.690919082729\\
0.772602739726027	160.836955848414\\
0.775342465753425	157.897666029183\\
0.778082191780822	153.919480689949\\
0.780821917808219	145.420694464997\\
0.783561643835616	144.035345094101\\
0.786301369863014	136.10099542163\\
0.789041095890411	134.473266619243\\
0.791780821917808	137.229052633758\\
0.794520547945206	137.488202076076\\
0.797260273972603	140.661959845158\\
0.8	148.97686593711\\
0.802739726027397	150.738764922326\\
0.805479452054795	148.55197604788\\
0.808219178082192	156.188997622138\\
0.810958904109589	154.813608997138\\
0.813698630136986	156.311460567806\\
0.816438356164384	152.888718709289\\
0.819178082191781	152.927653417231\\
0.821917808219178	152.078481599294\\
0.824657534246575	146.296145549605\\
0.827397260273973	147.233312070834\\
0.83013698630137	146.852504128957\\
0.832876712328767	148.477084548427\\
0.835616438356164	150.638302010418\\
0.838356164383562	155.786415107437\\
0.841095890410959	156.317981332918\\
0.843835616438356	157.467984462635\\
0.846575342465753	153.388578530689\\
0.849315068493151	149.679841071838\\
0.852054794520548	154.433637584727\\
0.854794520547945	148.588473748536\\
0.857534246575343	147.674579040045\\
0.86027397260274	153.140992850619\\
0.863013698630137	149.546695402947\\
0.865753424657534	155.544056229223\\
0.868493150684931	152.525174703832\\
0.871232876712329	152.495015327129\\
0.873972602739726	155.699278567445\\
0.876712328767123	151.312728535013\\
0.879452054794521	154.879046376174\\
0.882191780821918	155.73274324006\\
0.884931506849315	147.766794707856\\
0.887671232876712	145.650034665294\\
0.89041095890411	145.428196138604\\
0.893150684931507	145.039146139169\\
0.895890410958904	142.443962641482\\
0.898630136986301	141.447664444109\\
0.901369863013699	143.806524234872\\
0.904109589041096	140.844130535495\\
0.906849315068493	136.649176435065\\
0.90958904109589	138.094839675866\\
0.912328767123288	137.161668332778\\
0.915068493150685	135.324743092496\\
0.917808219178082	132.997475842429\\
0.920547945205479	133.991282670329\\
0.923287671232877	136.935003575642\\
0.926027397260274	134.301857765496\\
0.928767123287671	132.867441916953\\
0.931506849315068	129.910826878178\\
0.934246575342466	133.100713354667\\
0.936986301369863	135.140581303302\\
0.93972602739726	134.103150328445\\
0.942465753424658	132.566171158659\\
0.945205479452055	132.533026838085\\
0.947945205479452	136.503896077426\\
0.950684931506849	134.934786030036\\
0.953424657534247	137.945303459429\\
0.956164383561644	137.160951884794\\
0.958904109589041	138.936656063378\\
0.961643835616438	146.159664179304\\
0.964383561643836	150.801562825382\\
0.967123287671233	153.196330945489\\
0.96986301369863	157.036344154736\\
0.972602739726027	156.096047455391\\
0.975342465753425	157.502031626541\\
0.978082191780822	157.53941350086\\
0.980821917808219	161.339287819473\\
0.983561643835616	151.980862353629\\
0.986301369863014	160.793772981039\\
0.989041095890411	160.957066998124\\
0.991780821917808	157.723515759913\\
0.994520547945206	164.061736902001\\
0.997260273972603	166.019320922482\\
1	169.850593654253\\
};
\addlegendentry{Stock Price 1}

\addplot [color=black]
  table[row sep=crcr]{%
0	400\\
0.00273972602739726	407.96360559304\\
0.00547945205479452	420.572339109633\\
0.00821917808219178	397.734610843014\\
0.010958904109589	403.732043849234\\
0.0136986301369863	392.18447005154\\
0.0164383561643836	400.763433787269\\
0.0191780821917808	389.49276752464\\
0.0219178082191781	396.219213908716\\
0.0246575342465753	382.740964065784\\
0.0273972602739726	399.468598547083\\
0.0301369863013699	401.928972648101\\
0.0328767123287671	377.026248608405\\
0.0356164383561644	378.720272517523\\
0.0383561643835616	397.843304354504\\
0.0410958904109589	415.475357166764\\
0.0438356164383562	415.606240854761\\
0.0465753424657534	427.103214783991\\
0.0493150684931507	427.383970856742\\
0.0520547945205479	434.327238206359\\
0.0547945205479452	457.397925676899\\
0.0575342465753425	459.645621587719\\
0.0602739726027397	468.254807904623\\
0.063013698630137	465.928252411049\\
0.0657534246575343	458.948807710194\\
0.0684931506849315	466.741164324026\\
0.0712328767123288	462.447363200619\\
0.073972602739726	454.93990282007\\
0.0767123287671233	446.386202348454\\
0.0794520547945206	439.846447587605\\
0.0821917808219178	430.01404406185\\
0.0849315068493151	439.680998762416\\
0.0876712328767123	459.303864694203\\
0.0904109589041096	450.268767732807\\
0.0931506849315069	462.964322330741\\
0.0958904109589041	467.924371489055\\
0.0986301369863014	451.485489014765\\
0.101369863013699	459.454968764797\\
0.104109589041096	482.244407313587\\
0.106849315068493	487.602110937239\\
0.10958904109589	498.400918451368\\
0.112328767123288	469.016427451084\\
0.115068493150685	442.929518127753\\
0.117808219178082	449.093094576783\\
0.120547945205479	465.100590406336\\
0.123287671232877	475.951541082315\\
0.126027397260274	503.005531285706\\
0.128767123287671	503.179640261331\\
0.131506849315069	486.932263545198\\
0.134246575342466	497.017689963937\\
0.136986301369863	509.278279019477\\
0.13972602739726	516.04998795005\\
0.142465753424658	513.653714870725\\
0.145205479452055	545.827298730956\\
0.147945205479452	551.669272833105\\
0.150684931506849	536.20165381738\\
0.153424657534247	539.131679785536\\
0.156164383561644	538.329820251887\\
0.158904109589041	526.561640499816\\
0.161643835616438	533.93696309957\\
0.164383561643836	506.629920683876\\
0.167123287671233	493.615921139693\\
0.16986301369863	503.602060382775\\
0.172602739726027	485.136536172862\\
0.175342465753425	484.870547220197\\
0.178082191780822	482.244611522244\\
0.180821917808219	520.045439672491\\
0.183561643835616	526.42276785755\\
0.186301369863014	545.271228859515\\
0.189041095890411	549.045855421485\\
0.191780821917808	525.671081896218\\
0.194520547945205	516.10047864601\\
0.197260273972603	515.08607762107\\
0.2	525.178120278031\\
0.202739726027397	512.303564476509\\
0.205479452054795	531.280218144339\\
0.208219178082192	567.198910295177\\
0.210958904109589	559.901794531981\\
0.213698630136986	568.622891256361\\
0.216438356164384	588.201069120678\\
0.219178082191781	605.353279108878\\
0.221917808219178	612.99073567812\\
0.224657534246575	607.044142919408\\
0.227397260273973	608.493584921337\\
0.23013698630137	638.125829664841\\
0.232876712328767	645.289746125387\\
0.235616438356164	636.934233842402\\
0.238356164383562	638.960646432465\\
0.241095890410959	636.815800348304\\
0.243835616438356	620.030965788979\\
0.246575342465753	621.887505925732\\
0.249315068493151	649.927671204805\\
0.252054794520548	623.193499645884\\
0.254794520547945	614.969464753902\\
0.257534246575342	624.66047794701\\
0.26027397260274	624.662258742991\\
0.263013698630137	636.980431453306\\
0.265753424657534	649.281173235292\\
0.268493150684932	683.213778819587\\
0.271232876712329	684.30664782682\\
0.273972602739726	637.830922236214\\
0.276712328767123	639.052968047042\\
0.279452054794521	638.67926800511\\
0.282191780821918	644.212418223124\\
0.284931506849315	633.24093554618\\
0.287671232876712	615.129705246973\\
0.29041095890411	613.296092265726\\
0.293150684931507	584.226352319878\\
0.295890410958904	609.732193883247\\
0.298630136986301	618.012731152719\\
0.301369863013699	575.580652772599\\
0.304109589041096	560.898485186764\\
0.306849315068493	542.004253534119\\
0.30958904109589	550.518752065297\\
0.312328767123288	547.222343527422\\
0.315068493150685	557.43801354192\\
0.317808219178082	553.311573299177\\
0.320547945205479	544.292705015901\\
0.323287671232877	531.021334186133\\
0.326027397260274	525.60480757492\\
0.328767123287671	537.499766435726\\
0.331506849315069	531.016580607072\\
0.334246575342466	534.611135484567\\
0.336986301369863	526.800336433092\\
0.33972602739726	539.885192467727\\
0.342465753424658	543.424359647753\\
0.345205479452055	544.900084503933\\
0.347945205479452	539.68284598504\\
0.350684931506849	551.438004748468\\
0.353424657534247	553.210396703653\\
0.356164383561644	552.41963930564\\
0.358904109589041	547.90147347079\\
0.361643835616438	549.78486627231\\
0.364383561643836	549.270230782006\\
0.367123287671233	547.606392855334\\
0.36986301369863	545.312611472833\\
0.372602739726027	555.237134324303\\
0.375342465753425	529.254042733342\\
0.378082191780822	529.570358305877\\
0.380821917808219	533.520212332949\\
0.383561643835616	542.558595063971\\
0.386301369863014	574.180124405726\\
0.389041095890411	546.682138463533\\
0.391780821917808	543.61933315882\\
0.394520547945205	518.358412109003\\
0.397260273972603	513.335454002212\\
0.4	533.658231084881\\
0.402739726027397	553.94772046371\\
0.405479452054795	568.921203813728\\
0.408219178082192	543.65679297717\\
0.410958904109589	540.030830874086\\
0.413698630136986	539.724683692673\\
0.416438356164384	533.878862918113\\
0.419178082191781	511.763624669267\\
0.421917808219178	526.755738130458\\
0.424657534246575	530.430106738243\\
0.427397260273973	520.797408932687\\
0.43013698630137	542.392269401187\\
0.432876712328767	553.704990784296\\
0.435616438356164	552.646754954698\\
0.438356164383562	556.97187033313\\
0.441095890410959	547.881917262701\\
0.443835616438356	545.800959207026\\
0.446575342465753	539.89891229075\\
0.449315068493151	526.259198145372\\
0.452054794520548	523.481681244018\\
0.454794520547945	524.402193062985\\
0.457534246575342	526.355002251738\\
0.46027397260274	525.011225686429\\
0.463013698630137	547.379127677937\\
0.465753424657534	543.449980483713\\
0.468493150684932	549.461172254916\\
0.471232876712329	542.836518268561\\
0.473972602739726	543.903836697286\\
0.476712328767123	542.915919330966\\
0.479452054794521	529.853832975293\\
0.482191780821918	534.684972635545\\
0.484931506849315	557.684927978366\\
0.487671232876712	547.212614296039\\
0.49041095890411	570.916634821383\\
0.493150684931507	553.215550767128\\
0.495890410958904	534.657914207485\\
0.498630136986301	520.678347956711\\
0.501369863013699	515.825415119058\\
0.504109589041096	520.329905650539\\
0.506849315068493	494.397506927451\\
0.50958904109589	503.162540729066\\
0.512328767123288	525.237094223445\\
0.515068493150685	548.941620941395\\
0.517808219178082	520.264272044198\\
0.520547945205479	500.404916502013\\
0.523287671232877	520.929613905161\\
0.526027397260274	522.114494940477\\
0.528767123287671	519.305452409455\\
0.531506849315069	503.010218423948\\
0.534246575342466	509.254346335497\\
0.536986301369863	499.383765529021\\
0.53972602739726	497.861089734232\\
0.542465753424658	503.038304719954\\
0.545205479452055	505.974900149416\\
0.547945205479452	510.190005353326\\
0.550684931506849	512.056914433918\\
0.553424657534247	480.830060526845\\
0.556164383561644	488.179356683765\\
0.558904109589041	483.163540826743\\
0.561643835616438	519.160545363351\\
0.564383561643836	534.623485534612\\
0.567123287671233	508.662563388764\\
0.56986301369863	503.158202245898\\
0.572602739726027	502.422607516857\\
0.575342465753425	504.501811988186\\
0.578082191780822	498.654613881625\\
0.580821917808219	506.18217244197\\
0.583561643835616	501.35159148264\\
0.586301369863014	509.338427626331\\
0.589041095890411	502.380689648136\\
0.591780821917808	520.656572177782\\
0.594520547945205	515.492804443101\\
0.597260273972603	508.824741810772\\
0.6	500.819936083238\\
0.602739726027397	484.655968393687\\
0.605479452054795	477.156129595216\\
0.608219178082192	462.537655157078\\
0.610958904109589	462.410321132059\\
0.613698630136986	452.575928084868\\
0.616438356164384	457.188764381808\\
0.619178082191781	469.090447581929\\
0.621917808219178	481.93855355004\\
0.624657534246575	474.982344262858\\
0.627397260273973	473.872970781302\\
0.63013698630137	471.814224177657\\
0.632876712328767	471.102622963103\\
0.635616438356164	472.968932821499\\
0.638356164383562	458.448768879248\\
0.641095890410959	475.509195190687\\
0.643835616438356	477.794991825986\\
0.646575342465753	498.268277771367\\
0.649315068493151	500.455171505442\\
0.652054794520548	510.754729143847\\
0.654794520547945	468.616339735047\\
0.657534246575342	470.267928760305\\
0.66027397260274	494.012642158367\\
0.663013698630137	502.790861967801\\
0.665753424657534	482.608492501837\\
0.668493150684931	493.464485496496\\
0.671232876712329	501.10658606585\\
0.673972602739726	508.549054371053\\
0.676712328767123	518.438122680951\\
0.679452054794521	534.162998804515\\
0.682191780821918	516.788035634246\\
0.684931506849315	499.336019506695\\
0.687671232876712	506.508157714307\\
0.69041095890411	519.23995812259\\
0.693150684931507	514.297572561027\\
0.695890410958904	497.83959788681\\
0.698630136986301	493.093449383162\\
0.701369863013699	521.604761811476\\
0.704109589041096	510.341859117887\\
0.706849315068493	502.105286149618\\
0.70958904109589	477.291961557452\\
0.712328767123288	481.011099651368\\
0.715068493150685	483.554963196868\\
0.717808219178082	466.505392807579\\
0.720547945205479	477.03822596083\\
0.723287671232877	474.301633149776\\
0.726027397260274	454.757851685032\\
0.728767123287671	465.24828488778\\
0.731506849315068	457.946862096429\\
0.734246575342466	470.477105399055\\
0.736986301369863	467.003871908744\\
0.73972602739726	435.040810471877\\
0.742465753424657	439.64319387462\\
0.745205479452055	441.231941181508\\
0.747945205479452	446.580886526988\\
0.750684931506849	431.310594803637\\
0.753424657534247	413.511343696697\\
0.756164383561644	407.469476125714\\
0.758904109589041	418.255505457722\\
0.761643835616438	426.571204698016\\
0.764383561643836	420.669851857249\\
0.767123287671233	426.267913018921\\
0.76986301369863	410.220937528816\\
0.772602739726027	433.039592614417\\
0.775342465753425	426.965013251369\\
0.778082191780822	435.138353211446\\
0.780821917808219	430.486356787943\\
0.783561643835616	429.601637285531\\
0.786301369863014	434.679076929582\\
0.789041095890411	454.516167488781\\
0.791780821917808	465.784894424221\\
0.794520547945206	467.904223612053\\
0.797260273972603	468.141194529206\\
0.8	475.301515457996\\
0.802739726027397	485.146646319097\\
0.805479452054795	480.118008111485\\
0.808219178082192	487.246903748234\\
0.810958904109589	485.003185415271\\
0.813698630136986	482.20072978803\\
0.816438356164384	469.367063602809\\
0.819178082191781	469.272669830334\\
0.821917808219178	490.487192342317\\
0.824657534246575	490.127905696635\\
0.827397260273973	493.62335906553\\
0.83013698630137	510.12872455405\\
0.832876712328767	523.679084355812\\
0.835616438356164	535.236884851473\\
0.838356164383562	538.209189779121\\
0.841095890410959	543.691716104983\\
0.843835616438356	537.274444973974\\
0.846575342465753	532.330791131065\\
0.849315068493151	511.072855275571\\
0.852054794520548	519.770721938649\\
0.854794520547945	523.8005184525\\
0.857534246575343	527.74260166869\\
0.86027397260274	512.548851115281\\
0.863013698630137	502.101521507237\\
0.865753424657534	516.938885307805\\
0.868493150684931	514.585100298238\\
0.871232876712329	516.734380185597\\
0.873972602739726	512.517531644111\\
0.876712328767123	517.043457360642\\
0.879452054794521	507.112443824115\\
0.882191780821918	508.233556002713\\
0.884931506849315	500.521764107805\\
0.887671232876712	510.87582642367\\
0.89041095890411	520.886798148964\\
0.893150684931507	490.557785474306\\
0.895890410958904	477.878415249876\\
0.898630136986301	488.521374540632\\
0.901369863013699	466.848963599175\\
0.904109589041096	478.163758500616\\
0.906849315068493	495.031980387705\\
0.90958904109589	480.921486082337\\
0.912328767123288	493.190156684245\\
0.915068493150685	486.697609271476\\
0.917808219178082	467.546291616627\\
0.920547945205479	475.955907316709\\
0.923287671232877	492.342441985481\\
0.926027397260274	498.067343708804\\
0.928767123287671	497.748303860005\\
0.931506849315068	501.812701125412\\
0.934246575342466	507.604576997675\\
0.936986301369863	521.072447820014\\
0.93972602739726	532.371679945423\\
0.942465753424658	534.066654806878\\
0.945205479452055	527.730163756015\\
0.947945205479452	517.243584252445\\
0.950684931506849	517.462129941814\\
0.953424657534247	504.730190668915\\
0.956164383561644	501.867834697969\\
0.958904109589041	517.171109666928\\
0.961643835616438	529.021886600151\\
0.964383561643836	518.932574759321\\
0.967123287671233	513.310183986313\\
0.96986301369863	504.891414672069\\
0.972602739726027	505.438444744469\\
0.975342465753425	486.2434427195\\
0.978082191780822	476.164273970375\\
0.980821917808219	485.687219306579\\
0.983561643835616	472.876525557043\\
0.986301369863014	486.39050384\\
0.989041095890411	466.295049765334\\
0.991780821917808	465.518671163182\\
0.994520547945206	458.632849146597\\
0.997260273972603	443.383489287191\\
1	431.301069924698\\
};
\addlegendentry{Stock Price 2}

\end{axis}
\end{tikzpicture}%

%% file: tikz/21.tex
%
%
\definecolor{mycolor1}{rgb}{0.00000,0.44700,0.74100}%
\definecolor{mycolor2}{rgb}{0.85000,0.32500,0.09800}%
\begin{tikzpicture}

\begin{axis}[%
width=2.6in,
height=2.6in,
at={(0.758in,0.481in)},
scale only axis,
unbounded coords=jump,
xmin=0,
xmax=1,
xlabel style={font=\color{white!15!black}},
xlabel={Time (years)},
ymin=0,
ymax=1000,
ylabel style={font=\color{white!15!black}},
ylabel={Value (\$)},
axis background/.style={fill=white},
legend style={legend cell align=left, align=left, draw=white!15!black}
]
\addplot [color=mycolor1]
  table[row sep=crcr]{%
0	600\\
0.00273972602739726	610.456607436612\\
0.00547945205479452	625.207662065027\\
0.00821917808219178	604.942497412601\\
0.010958904109589	608.83587966996\\
0.0136986301369863	593.981593019923\\
0.0164383561643836	609.884876645812\\
0.0191780821917808	595.81146661082\\
0.0219178082191781	612.205746347628\\
0.0246575342465753	597.577067539528\\
0.0273972602739726	619.456875562484\\
0.0301369863013699	618.930293635824\\
0.0328767123287671	597.93440122224\\
0.0356164383561644	602.38983270871\\
0.0383561643835616	619.415714923604\\
0.0410958904109589	639.919909328201\\
0.0438356164383562	650.688880082302\\
0.0465753424657534	653.844005792905\\
0.0493150684931507	654.780823263363\\
0.0520547945205479	673.801275427572\\
0.0547945205479452	693.428715743612\\
0.0575342465753425	698.105682930401\\
0.0602739726027397	703.979401161119\\
0.063013698630137	685.15539462277\\
0.0657534246575343	680.629963579636\\
0.0684931506849315	679.761087680673\\
0.0712328767123288	668.985811013184\\
0.073972602739726	662.288075712746\\
0.0767123287671233	658.803816349358\\
0.0794520547945206	659.302708088936\\
0.0821917808219178	647.828036850403\\
0.0849315068493151	653.341080899555\\
0.0876712328767123	665.566078536799\\
0.0904109589041096	649.9858440073\\
0.0931506849315069	667.176203709912\\
0.0958904109589041	681.042760772705\\
0.0986301369863014	662.940319425361\\
0.101369863013699	671.223960421398\\
0.104109589041096	696.111492234015\\
0.106849315068493	714.727249822916\\
0.10958904109589	732.483108888314\\
0.112328767123288	701.775391247701\\
0.115068493150685	673.40344510033\\
0.117808219178082	676.305770627099\\
0.120547945205479	691.07405783019\\
0.123287671232877	683.783327034439\\
0.126027397260274	702.861021289968\\
0.128767123287671	703.167766218016\\
0.131506849315069	694.019804524313\\
0.134246575342466	695.897346569655\\
0.136986301369863	721.016876428544\\
0.13972602739726	730.88149614512\\
0.142465753424658	726.776595747375\\
0.145205479452055	759.964463036085\\
0.147945205479452	756.193120379147\\
0.150684931506849	742.463441901544\\
0.153424657534247	747.172174591729\\
0.156164383561644	742.153001089385\\
0.158904109589041	723.55132030747\\
0.161643835616438	732.217071214887\\
0.164383561643836	699.367030227346\\
0.167123287671233	683.246243808421\\
0.16986301369863	695.811669869767\\
0.172602739726027	675.986596486072\\
0.175342465753425	674.271125721239\\
0.178082191780822	671.517095850835\\
0.180821917808219	705.096273939093\\
0.183561643835616	711.959710839833\\
0.186301369863014	731.015399428858\\
0.189041095890411	735.316057473785\\
0.191780821917808	719.731651812739\\
0.194520547945205	709.719173371867\\
0.197260273972603	708.117708439119\\
0.2	716.393998463042\\
0.202739726027397	700.019287082522\\
0.205479452054795	727.897876217253\\
0.208219178082192	769.782678256871\\
0.210958904109589	772.708037348469\\
0.213698630136986	786.237911222647\\
0.216438356164384	805.776328165463\\
0.219178082191781	828.912452494343\\
0.221917808219178	829.884915630872\\
0.224657534246575	816.587107353087\\
0.227397260273973	815.631943685936\\
0.23013698630137	841.779546426953\\
0.232876712328767	842.796152856607\\
0.235616438356164	838.220113321923\\
0.238356164383562	842.176140899254\\
0.241095890410959	831.145774268466\\
0.243835616438356	824.827304542269\\
0.246575342465753	819.030461485072\\
0.249315068493151	844.361841840822\\
0.252054794520548	815.211706249563\\
0.254794520547945	809.087159632022\\
0.257534246575342	808.402031005671\\
0.26027397260274	808.009607652216\\
0.263013698630137	815.439640274172\\
0.265753424657534	827.34699518455\\
0.268493150684932	855.909622694812\\
0.271232876712329	854.971920242574\\
0.273972602739726	805.01721360439\\
0.276712328767123	817.954159043698\\
0.279452054794521	812.735913682938\\
0.282191780821918	816.424846968858\\
0.284931506849315	801.904329500146\\
0.287671232876712	780.935827493022\\
0.29041095890411	774.999242112266\\
0.293150684931507	750.49128601912\\
0.295890410958904	778.978885870711\\
0.298630136986301	784.561116695933\\
0.301369863013699	741.231637058712\\
0.304109589041096	727.988465898067\\
0.306849315068493	705.986479986268\\
0.30958904109589	714.76562324698\\
0.312328767123288	706.819013894249\\
0.315068493150685	714.871513595521\\
0.317808219178082	712.799461818524\\
0.320547945205479	704.890544060417\\
0.323287671232877	693.165231532542\\
0.326027397260274	690.313161333766\\
0.328767123287671	702.006883758167\\
0.331506849315069	695.661212791602\\
0.334246575342466	699.803733658727\\
0.336986301369863	693.426031999717\\
0.33972602739726	702.446616941825\\
0.342465753424658	704.273025082501\\
0.345205479452055	697.635196282761\\
0.347945205479452	698.317312125258\\
0.350684931506849	709.727274090555\\
0.353424657534247	716.897144195485\\
0.356164383561644	721.452951605226\\
0.358904109589041	711.043338925168\\
0.361643835616438	714.124329159313\\
0.364383561643836	713.551986435589\\
0.367123287671233	714.351343999127\\
0.36986301369863	702.806206017451\\
0.372602739726027	710.407682562387\\
0.375342465753425	682.287516463501\\
0.378082191780822	680.094266176405\\
0.380821917808219	690.766256034329\\
0.383561643835616	700.921530955275\\
0.386301369863014	738.260547098107\\
0.389041095890411	709.109450864804\\
0.391780821917808	705.694569299175\\
0.394520547945205	678.835995605444\\
0.397260273972603	667.568883576025\\
0.4	683.757384692775\\
0.402739726027397	709.788113582959\\
0.405479452054795	728.06647060208\\
0.408219178082192	708.659091231024\\
0.410958904109589	714.480701025409\\
0.413698630136986	710.697994320582\\
0.416438356164384	705.351971209\\
0.419178082191781	684.614777540354\\
0.421917808219178	698.522994394201\\
0.424657534246575	700.572114832238\\
0.427397260273973	688.20199159183\\
0.43013698630137	708.592591318172\\
0.432876712328767	719.560082979751\\
0.435616438356164	716.30610078881\\
0.438356164383562	714.80195412978\\
0.441095890410959	711.271966014618\\
0.443835616438356	714.033970128724\\
0.446575342465753	713.587068376002\\
0.449315068493151	704.133325955199\\
0.452054794520548	697.88987403771\\
0.454794520547945	702.918813141354\\
0.457534246575342	702.758555279582\\
0.46027397260274	694.106438208963\\
0.463013698630137	718.63931838309\\
0.465753424657534	714.583720246824\\
0.468493150684932	723.328769693984\\
0.471232876712329	724.713453673677\\
0.473972602739726	734.987638888809\\
0.476712328767123	738.298193326426\\
0.479452054794521	723.765771704657\\
0.482191780821918	719.432371262633\\
0.484931506849315	743.279045048278\\
0.487671232876712	734.349458839359\\
0.49041095890411	759.881100745252\\
0.493150684931507	740.228120552183\\
0.495890410958904	722.245360240747\\
0.498630136986301	710.184368858405\\
0.501369863013699	699.228429975757\\
0.504109589041096	705.169887027034\\
0.506849315068493	676.873797841383\\
0.50958904109589	687.871352311866\\
0.512328767123288	714.604702066001\\
0.515068493150685	743.293834111368\\
0.517808219178082	703.908180665067\\
0.520547945205479	677.27967531361\\
0.523287671232877	701.166223157158\\
0.526027397260274	699.303911590304\\
0.528767123287671	700.283843153784\\
0.531506849315069	682.151746512358\\
0.534246575342466	690.397970020362\\
0.536986301369863	672.582959159473\\
0.53972602739726	672.698702281621\\
0.542465753424658	677.132503184176\\
0.545205479452055	677.963341142038\\
0.547945205479452	680.527736241497\\
0.550684931506849	685.171506595045\\
0.553424657534247	658.022960190915\\
0.556164383561644	669.469428570092\\
0.558904109589041	665.358836807813\\
0.561643835616438	706.336664712231\\
0.564383561643836	723.591985833909\\
0.567123287671233	695.475902978343\\
0.56986301369863	691.700844497721\\
0.572602739726027	692.238658186973\\
0.575342465753425	687.692222066703\\
0.578082191780822	683.50279228145\\
0.580821917808219	693.551848393979\\
0.583561643835616	678.705386036346\\
0.586301369863014	679.153410400386\\
0.589041095890411	674.365425371034\\
0.591780821917808	687.801482278767\\
0.594520547945205	677.584089337725\\
0.597260273972603	675.172258593801\\
0.6	674.908990952581\\
0.602739726027397	658.495794683955\\
0.605479452054795	654.260892181313\\
0.608219178082192	634.927593703864\\
0.610958904109589	633.245515861626\\
0.613698630136986	627.537850230244\\
0.616438356164384	626.948288123391\\
0.619178082191781	636.664640402105\\
0.621917808219178	649.245838222339\\
0.624657534246575	644.697760506719\\
0.627397260273973	651.800907155106\\
0.63013698630137	648.448760738226\\
0.632876712328767	654.655065183738\\
0.635616438356164	661.507004334791\\
0.638356164383562	655.416013245354\\
0.641095890410959	674.082070275002\\
0.643835616438356	679.114341144448\\
0.646575342465753	701.362271992831\\
0.649315068493151	695.096413339844\\
0.652054794520548	708.364610364432\\
0.654794520547945	672.630015892168\\
0.657534246575342	679.884641086507\\
0.66027397260274	698.20943731031\\
0.663013698630137	708.794649206092\\
0.665753424657534	693.200843191742\\
0.668493150684931	695.75983135778\\
0.671232876712329	707.493737109587\\
0.673972602739726	713.232097351725\\
0.676712328767123	723.7122793792\\
0.679452054794521	739.089132782406\\
0.682191780821918	718.801608049227\\
0.684931506849315	702.772428810292\\
0.687671232876712	707.293794540685\\
0.69041095890411	715.518530751803\\
0.693150684931507	707.340787162963\\
0.695890410958904	687.503655712557\\
0.698630136986301	678.607296452761\\
0.701369863013699	700.746671534677\\
0.704109589041096	700.867217708716\\
0.706849315068493	692.640325044095\\
0.70958904109589	662.33201018187\\
0.712328767123288	671.730797568909\\
0.715068493150685	673.607593485049\\
0.717808219178082	649.392537409363\\
0.720547945205479	658.23442499284\\
0.723287671232877	651.263949620937\\
0.726027397260274	629.923773374804\\
0.728767123287671	641.62473182401\\
0.731506849315068	635.635245736475\\
0.734246575342466	647.006314048377\\
0.736986301369863	640.583213378223\\
0.73972602739726	605.970386552303\\
0.742465753424657	608.866356083803\\
0.745205479452055	607.47456173017\\
0.747945205479452	615.596021770157\\
0.750684931506849	594.889809379115\\
0.753424657534247	583.635444001938\\
0.756164383561644	579.216480268645\\
0.758904109589041	581.925601779033\\
0.761643835616438	585.538189908565\\
0.764383561643836	585.937433749811\\
0.767123287671233	593.667454670196\\
0.76986301369863	570.911856611545\\
0.772602739726027	593.876548462831\\
0.775342465753425	584.862679280552\\
0.778082191780822	589.057833901395\\
0.780821917808219	575.90705125294\\
0.783561643835616	573.636982379631\\
0.786301369863014	570.780072351212\\
0.789041095890411	588.989434108024\\
0.791780821917808	603.013947057979\\
0.794520547945206	605.392425688129\\
0.797260273972603	608.803154374364\\
0.8	624.278381395105\\
0.802739726027397	635.885411241422\\
0.805479452054795	628.669984159365\\
0.808219178082192	643.435901370371\\
0.810958904109589	639.816794412409\\
0.813698630136986	638.512190355836\\
0.816438356164384	622.255782312098\\
0.819178082191781	622.200323247566\\
0.821917808219178	642.565673941611\\
0.824657534246575	636.42405124624\\
0.827397260273973	640.856671136364\\
0.83013698630137	656.981228683007\\
0.832876712328767	672.156168904239\\
0.835616438356164	685.875186861891\\
0.838356164383562	693.995604886558\\
0.841095890410959	700.009697437901\\
0.843835616438356	694.74242943661\\
0.846575342465753	685.719369661754\\
0.849315068493151	660.752696347408\\
0.852054794520548	674.204359523377\\
0.854794520547945	672.388992201036\\
0.857534246575343	675.417180708736\\
0.86027397260274	665.689843965901\\
0.863013698630137	651.648216910184\\
0.865753424657534	672.482941537028\\
0.868493150684931	667.11027500207\\
0.871232876712329	669.229395512727\\
0.873972602739726	668.216810211556\\
0.876712328767123	668.356185895655\\
0.879452054794521	661.991490200289\\
0.882191780821918	663.966299242773\\
0.884931506849315	648.288558815661\\
0.887671232876712	656.525861088964\\
0.89041095890411	666.314994287569\\
0.893150684931507	635.596931613474\\
0.895890410958904	620.322377891358\\
0.898630136986301	629.969038984741\\
0.901369863013699	610.655487834047\\
0.904109589041096	619.00788903611\\
0.906849315068493	631.68115682277\\
0.90958904109589	619.016325758204\\
0.912328767123288	630.351825017023\\
0.915068493150685	622.022352363973\\
0.917808219178082	600.543767459056\\
0.920547945205479	609.947189987039\\
0.923287671232877	629.277445561123\\
0.926027397260274	632.3692014743\\
0.928767123287671	630.615745776958\\
0.931506849315068	631.72352800359\\
0.934246575342466	640.705290352343\\
0.936986301369863	656.213029123316\\
0.93972602739726	666.474830273868\\
0.942465753424658	666.632825965537\\
0.945205479452055	660.2631905941\\
0.947945205479452	653.747480329871\\
0.950684931506849	652.39691597185\\
0.953424657534247	642.675494128344\\
0.956164383561644	639.028786582763\\
0.958904109589041	656.107765730305\\
0.961643835616438	675.181550779455\\
0.964383561643836	669.734137584703\\
0.967123287671233	666.506514931802\\
0.96986301369863	661.927758826805\\
0.972602739726027	661.53449219986\\
0.975342465753425	643.745474346041\\
0.978082191780822	633.703687471235\\
0.980821917808219	647.026507126052\\
0.983561643835616	624.857387910672\\
0.986301369863014	647.184276821039\\
0.989041095890411	627.252116763459\\
0.991780821917808	623.242186923095\\
0.994520547945206	622.694586048599\\
0.997260273972603	609.402810209673\\
1	601.151663578951\\
};
\addlegendentry{Sum of Stock Prices}

\addplot [color=mycolor2]
  table[row sep=crcr]{%
0	40\\
0.00273972602739726	41.5349170151103\\
0.00547945205479452	43.1099348430366\\
0.00821917808219178	44.6297728355705\\
0.010958904109589	46.1602780711987\\
0.0136986301369863	47.6501151393872\\
0.0164383561643836	49.1836827933936\\
0.0191780821917808	50.6785021993912\\
0.0219178082191781	52.2186630058312\\
0.0246575342465753	53.7183145340934\\
0.0273972602739726	55.2788196536839\\
0.0301369863013699	56.8378941739416\\
0.0328767123287671	58.338306769361\\
0.0356164383561644	59.8512516455768\\
0.0383561643835616	61.4121166985622\\
0.0410958904109589	63.0308538879697\\
0.0438356164383562	64.6800962016486\\
0.0465753424657534	66.3383335635815\\
0.0493150684931507	67.9992814361829\\
0.0520547945205479	69.7145582075578\\
0.0547945205479452	71.4860640462805\\
0.0575342465753425	73.2710455733897\\
0.0602739726027397	75.0729902869957\\
0.063013698630137	76.8206183100293\\
0.0657534246575343	78.5551885685111\\
0.0684931506849315	80.287287264291\\
0.0712328767123288	81.9880365019555\\
0.073972602739726	83.6692618676962\\
0.0767123287671233	85.3403273445185\\
0.0794520547945206	87.0129177684011\\
0.0821917808219178	88.6517154176234\\
0.0849315068493151	90.3068869659419\\
0.0876712328767123	91.9984066614156\\
0.0904109589041096	93.6435993516491\\
0.0931506849315069	95.3401969480009\\
0.0958904109589041	97.0784021256714\\
0.0986301369863014	98.7622704552432\\
0.101369863013699	100.471177682494\\
0.104109589041096	102.255414645692\\
0.106849315068493	104.096193027018\\
0.10958904109589	105.991075177627\\
0.112328767123288	107.792286013941\\
0.115068493150685	109.506682824587\\
0.117808219178082	111.230070002506\\
0.120547945205479	112.999047390912\\
0.123287671232877	114.745559637455\\
0.126027397260274	116.551322948691\\
0.128767123287671	118.358119491116\\
0.131506849315069	120.136438541639\\
0.134246575342466	121.92071889108\\
0.136986301369863	123.784007007708\\
0.13972602739726	125.67847422037\\
0.142465753424658	127.560044523742\\
0.145205479452055	129.54700192425\\
0.147945205479452	131.522041530421\\
0.150684931506849	133.453320281723\\
0.153424657534247	135.39977760477\\
0.156164383561644	137.330189884859\\
0.158904109589041	139.200692080673\\
0.161643835616438	141.099334430578\\
0.164383561643836	142.891399274609\\
0.167123287671233	144.631036379705\\
0.16986301369863	146.41184940158\\
0.172602739726027	148.127731256341\\
0.175342465753425	149.838067026335\\
0.178082191780822	151.539408815716\\
0.180821917808219	153.352126817825\\
0.183561643835616	155.18777015193\\
0.186301369863014	157.087098313054\\
0.189041095890411	159.000931329293\\
0.191780821917808	160.862514348476\\
0.194520547945205	162.690451444931\\
0.197260273972603	164.513080720951\\
0.2	166.363917396167\\
0.202739726027397	168.159083085842\\
0.205479452054795	170.049669462649\\
0.208219178082192	172.084064914823\\
0.210958904109589	174.12864814737\\
0.213698630136986	176.220095456616\\
0.216438356164384	178.379408486365\\
0.219178082191781	180.619349867485\\
0.221917808219178	182.862809322996\\
0.224657534246575	185.059786352798\\
0.227397260273973	187.253528672252\\
0.23013698630137	189.539703150379\\
0.232876712328767	191.829607516092\\
0.235616438356164	194.103369398098\\
0.238356164383562	196.391381245938\\
0.241095890410959	198.640014430137\\
0.243835616438356	200.866065311267\\
0.246575342465753	203.071333693588\\
0.249315068493151	205.368488057455\\
0.252054794520548	207.559807805023\\
0.254794520547945	209.728919168301\\
0.257534246575342	211.895662786589\\
0.26027397260274	214.061106806971\\
0.263013698630137	216.254115145305\\
0.265753424657534	218.491378966904\\
0.268493150684932	220.834999935188\\
0.271232876712329	223.175269943219\\
0.273972602739726	225.32851088426\\
0.276712328767123	227.530565168086\\
0.279452054794521	229.713069416769\\
0.282191780821918	231.909712000324\\
0.284931506849315	234.051256112687\\
0.287671232876712	236.11285501491\\
0.29041095890411	238.151845745376\\
0.293150684931507	240.096635960357\\
0.295890410958904	242.151705487895\\
0.298630136986301	244.228601916296\\
0.301369863013699	246.136842784602\\
0.304109589041096	247.993445714322\\
0.306849315068493	249.763804403678\\
0.30958904109589	251.568950282614\\
0.312328767123288	253.342803926497\\
0.315068493150685	255.148839464876\\
0.317808219178082	256.946776888771\\
0.320547945205479	258.713190826112\\
0.323287671232877	260.432594390497\\
0.326027397260274	262.140649260213\\
0.328767123287671	263.896328629103\\
0.331506849315069	265.626331698807\\
0.334246575342466	267.37346273464\\
0.336986301369863	269.094571702999\\
0.33972602739726	270.853079193828\\
0.342465753424658	272.619336780092\\
0.345205479452055	274.358160320096\\
0.347945205479452	276.10001937983\\
0.350684931506849	277.889944471901\\
0.353424657534247	279.710273240716\\
0.356164383561644	281.550073204421\\
0.358904109589041	283.345814567689\\
0.361643835616438	285.154900803962\\
0.364383561643836	286.961725552676\\
0.367123287671233	288.772186705664\\
0.36986301369863	290.532903388499\\
0.372602739726027	292.326842509641\\
0.375342465753425	293.998263151182\\
0.378082191780822	295.660268437604\\
0.380821917808219	297.36946278977\\
0.383561643835616	299.123773244572\\
0.386301369863014	301.044176103111\\
0.389041095890411	302.834682103607\\
0.391780821917808	304.610082781674\\
0.394520547945205	306.264723598758\\
0.397260273972603	307.868593311377\\
0.4	309.546253890151\\
0.402739726027397	311.342991717814\\
0.405479452054795	313.223795895949\\
0.408219178082192	315.015359454352\\
0.410958904109589	316.83409481191\\
0.413698630136986	318.635444361981\\
0.416438356164384	320.412017684613\\
0.419178082191781	322.091403721491\\
0.421917808219178	323.836647736357\\
0.424657534246575	325.591827936763\\
0.427397260273973	327.28828255938\\
0.43013698630137	329.082594321778\\
0.432876712328767	330.929900050673\\
0.435616438356164	332.761695603026\\
0.438356164383562	334.586407249666\\
0.441095890410959	336.39410771469\\
0.443835616438356	338.215589573672\\
0.446575342465753	340.03509584961\\
0.449315068493151	341.807973060566\\
0.452054794520548	343.549978516545\\
0.454794520547945	345.317395468783\\
0.457534246575342	347.084239215721\\
0.46027397260274	348.80755342651\\
0.463013698630137	350.655833750752\\
0.465753424657534	352.483624768679\\
0.468493150684932	354.35657793888\\
0.471232876712329	356.236920896008\\
0.473972602739726	358.170838248622\\
0.476712328767123	360.122273513266\\
0.479452054794521	361.997712992888\\
0.482191780821918	363.850542152801\\
0.484931506849315	365.830061640642\\
0.487671232876712	367.76222484871\\
0.49041095890411	369.831482590424\\
0.493150684931507	371.795079090319\\
0.495890410958904	373.661485861628\\
0.498630136986301	375.462431867387\\
0.501369863013699	377.203608656947\\
0.504109589041096	378.977772118011\\
0.506849315068493	380.595434313695\\
0.50958904109589	382.27462172376\\
0.512328767123288	384.103843354407\\
0.515068493150685	386.094972436224\\
0.517808219178082	387.863175425163\\
0.520547945205479	389.479876271865\\
0.523287671232877	391.233764891291\\
0.526027397260274	392.977180369181\\
0.528767123287671	394.726544335024\\
0.531506849315069	396.370400700046\\
0.534246575342466	398.062914307105\\
0.536986301369863	399.650534382127\\
0.53972602739726	401.239111768023\\
0.542465753424658	402.854448493931\\
0.545205479452055	404.475051023497\\
0.547945205479452	406.111433786198\\
0.550684931506849	407.776344216477\\
0.553424657534247	409.275335277775\\
0.556164383561644	410.845109613294\\
0.558904109589041	412.38968329824\\
0.561643835616438	414.19012026632\\
0.564383561643836	416.099141762153\\
0.567123287671233	417.83084773939\\
0.56986301369863	419.53883790877\\
0.572602739726027	421.250552419583\\
0.575342465753425	422.933274799621\\
0.578082191780822	424.589128998507\\
0.580821917808219	426.310831206011\\
0.583561643835616	427.935320827388\\
0.586301369863014	429.563061418861\\
0.589041095890411	431.159226840837\\
0.591780821917808	432.845707361062\\
0.594520547945205	434.463557189559\\
0.597260273972603	436.065319899793\\
0.6	437.665576831748\\
0.602739726027397	439.153109630988\\
0.605479452054795	440.611574478252\\
0.608219178082192	441.935338953449\\
0.610958904109589	443.247573491298\\
0.613698630136986	444.519684825725\\
0.616438356164384	445.787891120839\\
0.619178082191781	447.126205902984\\
0.621917808219178	448.555870454977\\
0.624657534246575	449.952683192102\\
0.627397260273973	451.401963161882\\
0.63013698630137	452.826747865243\\
0.632876712328767	454.298098682734\\
0.635616438356164	455.821214554653\\
0.638356164383562	457.298547017094\\
0.641095890410959	458.918519385545\\
0.643835616438356	460.577469761857\\
0.646575342465753	462.40901469329\\
0.649315068493151	464.191969992252\\
0.652054794520548	466.07960913383\\
0.654794520547945	467.684228776639\\
0.657534246575342	469.347147026454\\
0.66027397260274	471.158028732196\\
0.663013698630137	473.055211585971\\
0.665753424657534	474.825004954397\\
0.668493150684931	476.616250291481\\
0.671232876712329	478.505522069339\\
0.673972602739726	480.443301403949\\
0.676712328767123	482.470154232901\\
0.679452054794521	484.628664646899\\
0.682191780821918	486.612731094748\\
0.684931506849315	488.457834997076\\
0.687671232876712	490.342904569082\\
0.69041095890411	492.301045893296\\
0.693150684931507	494.186547291934\\
0.695890410958904	495.893770574851\\
0.698630136986301	497.520496531751\\
0.701369863013699	499.350567458359\\
0.704109589041096	501.182091341073\\
0.706849315068493	502.93710283264\\
0.70958904109589	504.406653176265\\
0.712328767123288	505.966019025787\\
0.715068493150685	507.543765657503\\
0.717808219178082	508.886845033912\\
0.720547945205479	510.316923094655\\
0.723287671232877	511.678353812021\\
0.726027397260274	512.826793240257\\
0.728767123287671	514.093737359319\\
0.731506849315068	515.299927406999\\
0.734246575342466	516.62366526665\\
0.736986301369863	517.880859249853\\
0.73972602739726	518.774128222612\\
0.742465753424657	519.69855051775\\
0.745205479452055	520.608360634044\\
0.747945205479452	521.606787633864\\
0.750684931506849	522.378051300541\\
0.753424657534247	523.024630597175\\
0.756164383561644	523.621915408198\\
0.758904109589041	524.250337888883\\
0.761643835616438	524.920637276214\\
0.764383561643836	525.595933593827\\
0.767123287671233	526.362527179777\\
0.76986301369863	526.858568895937\\
0.772602739726027	527.631662402302\\
0.775342465753425	528.29517997396\\
0.778082191780822	529.010850782406\\
0.780821917808219	529.56247750505\\
0.783561643835616	530.085723494779\\
0.786301369863014	530.572695513264\\
0.789041095890411	531.296547705624\\
0.791780821917808	532.20532356217\\
0.794520547945206	533.146177981745\\
0.797260273972603	534.133492965863\\
0.8	535.333205259239\\
0.802739726027397	536.694526366757\\
0.805479452054795	537.954558446387\\
0.808219178082192	539.42595263619\\
0.810958904109589	540.845245409873\\
0.813698630136986	542.245712414847\\
0.816438356164384	543.403835960866\\
0.819178082191781	544.561486035287\\
0.821917808219178	546.032925583078\\
0.824657534246575	547.408736519943\\
0.827397260273973	548.855302699436\\
0.83013698630137	550.562414394595\\
0.832876712328767	552.518767807742\\
0.835616438356164	554.70424005408\\
0.838356164383562	557.027780431931\\
0.841095890410959	559.455434325165\\
0.843835616438356	561.791015787146\\
0.846575342465753	563.965775136421\\
0.849315068493151	565.686757400171\\
0.852054794520548	567.65734968078\\
0.854794520547945	569.59405590343\\
0.857534246575343	571.589411452566\\
0.86027397260274	573.394320105217\\
0.863013698630137	574.918631754309\\
0.865753424657534	576.868764855123\\
0.868493150684931	578.707294125255\\
0.871232876712329	580.59132786328\\
0.873972602739726	582.453728242951\\
0.876712328767123	584.319621176124\\
0.879452054794521	586.041169281796\\
0.882191780821918	587.809067274002\\
0.884931506849315	589.203850083001\\
0.887671232876712	590.800067887125\\
0.89041095890411	592.641570247213\\
0.893150684931507	593.695317669204\\
0.895890410958904	594.347239103993\\
0.898630136986301	595.260456782895\\
0.901369863013699	595.63722626733\\
0.904109589041096	596.253201453111\\
0.906849315068493	597.242580547347\\
0.90958904109589	597.848311872149\\
0.912328767123288	598.808931824513\\
0.915068493150685	599.501071795668\\
0.917808219178082	599.477147836357\\
0.920547945205479	599.7781206969\\
0.923287671232877	600.770371678835\\
0.926027397260274	601.877622738847\\
0.928767123287671	602.91778947645\\
0.931506849315068	604.002708118209\\
0.934246575342466	605.462558322923\\
0.936986301369863	607.597581274236\\
0.93972602739726	610.199818293208\\
0.942465753424658	612.809996948837\\
0.945205479452055	615.101860300261\\
0.947945205479452	617.050939035752\\
0.950684931506849	618.925344310457\\
0.953424657534247	620.227867772443\\
0.956164383561644	621.302710173904\\
0.958904109589041	623.517478257327\\
0.961643835616438	627.09617093721\\
0.964383561643836	630.255918895531\\
0.967123287671233	633.146907271599\\
0.96986301369863	635.621736225424\\
0.972602739726027	638.057636016706\\
0.975342465753425	638.515802112902\\
0.978082191780822	637.718153136228\\
0.980821917808219	638.825749377502\\
0.983561643835616	636.23590084884\\
0.986301369863014	638.115519013553\\
0.989041095890411	635.008439893326\\
0.991780821917808	630.564048765286\\
0.994520547945206	625.84605816631\\
0.997260273972603	607.825786777691\\
1	nan\\
};
\addlegendentry{Accumulated Cash}

\end{axis}
\end{tikzpicture}%

%% file: tikz/bls.tex
%
%
\definecolor{mycolor1}{rgb}{0.00000,0.44700,0.74100}%
\definecolor{mycolor2}{rgb}{0.85000,0.32500,0.09800}%
\begin{tikzpicture}

\begin{axis}[%
width=4.521in,
height=2.7in,
at={(0.758in,0.481in)},
scale only axis,
unbounded coords=jump,
xmin=0,
xmax=2,
xlabel style={font=\color{white!15!black}},
xlabel={Time (years)},
ymin=0,
ymax=35,
ylabel style={font=\color{white!15!black}},
ylabel={Value (\$)},
axis background/.style={fill=white}
]
\addplot [color=mycolor1, forget plot]
  table[row sep=crcr]{%
0	23.408520186585\\
0.00547945205479452	23.1035736432632\\
0.010958904109589	23.0382124145132\\
0.0164383561643836	22.8226487237875\\
0.0219178082191781	22.7556062607683\\
0.0273972602739726	22.6478355535167\\
0.0328767123287671	22.5533597367497\\
0.0383561643835616	22.4384012039514\\
0.0438356164383562	22.7172917495293\\
0.0493150684931507	22.8591933533265\\
0.0547945205479452	22.8731159930574\\
0.0602739726027397	22.5275461564523\\
0.0657534246575343	22.1579385161355\\
0.0712328767123288	21.7290035768066\\
0.0767123287671233	22.2755111903347\\
0.0821917808219178	22.4916119434092\\
0.0876712328767123	22.549980427105\\
0.0931506849315069	22.327732417992\\
0.0986301369863014	22.0723367242371\\
0.104109589041096	22.1913541763978\\
0.10958904109589	22.5364223816894\\
0.115068493150685	22.1510531198488\\
0.120547945205479	22.0526930627854\\
0.126027397260274	22.0344294116147\\
0.131506849315069	21.6053115102786\\
0.136986301369863	21.7788343537609\\
0.142465753424658	21.6917719345275\\
0.147945205479452	22.0909792149074\\
0.153424657534247	22.0059558202159\\
0.158904109589041	22.0942502420564\\
0.164383561643836	21.8459800531162\\
0.16986301369863	21.7137826804353\\
0.175342465753425	21.4705614614608\\
0.180821917808219	21.5502172072896\\
0.186301369863014	21.8173937237922\\
0.191780821917808	21.6921743368158\\
0.197260273972603	21.7830925027042\\
0.202739726027397	21.6770879876697\\
0.208219178082192	21.6983142827364\\
0.213698630136986	21.9481212025345\\
0.219178082191781	21.9739137105326\\
0.224657534246575	21.7513883878566\\
0.23013698630137	21.5719713346816\\
0.235616438356164	21.5370980099016\\
0.241095890410959	21.4979118757443\\
0.246575342465753	21.3219911966476\\
0.252054794520548	21.3580128058673\\
0.257534246575342	21.6122729823793\\
0.263013698630137	21.560453932254\\
0.268493150684932	21.5184986015236\\
0.273972602739726	21.0714751772585\\
0.279452054794521	20.9222218177979\\
0.284931506849315	20.659550327662\\
0.29041095890411	20.5890779249492\\
0.295890410958904	20.6843465451388\\
0.301369863013699	20.9650651655077\\
0.306849315068493	21.309357452105\\
0.312328767123288	21.2860923652053\\
0.317808219178082	21.2225680535645\\
0.323287671232877	21.3839277004219\\
0.328767123287671	21.482544117865\\
0.334246575342466	21.6384323872676\\
0.33972602739726	21.7999058603299\\
0.345205479452055	21.4883263891633\\
0.350684931506849	21.5984525989201\\
0.356164383561644	21.5795747354232\\
0.361643835616438	21.4233370729021\\
0.367123287671233	21.2834077260052\\
0.372602739726027	21.5386643070759\\
0.378082191780822	21.3730443766995\\
0.383561643835616	21.1253247533689\\
0.389041095890411	21.1671696851443\\
0.394520547945205	21.5626817378587\\
0.4	21.5679333782733\\
0.405479452054795	21.6313686903236\\
0.410958904109589	21.4373285372892\\
0.416438356164384	21.7774046317704\\
0.421917808219178	21.8037442023513\\
0.427397260273973	21.9158643740132\\
0.432876712328767	21.7124142042595\\
0.438356164383562	21.8913091150026\\
0.443835616438356	21.974420806867\\
0.449315068493151	21.7258123645765\\
0.454794520547945	21.7341406511631\\
0.46027397260274	21.6396887647536\\
0.465753424657534	21.6623656285409\\
0.471232876712329	21.6102492837979\\
0.476712328767123	21.5709780370555\\
0.482191780821918	21.3586725828152\\
0.487671232876712	21.2937840658047\\
0.493150684931507	21.4467760691642\\
0.498630136986301	21.801966618603\\
0.504109589041096	21.556986350778\\
0.50958904109589	22.0476115336629\\
0.515068493150685	22.3797391227374\\
0.520547945205479	22.4176632433331\\
0.526027397260274	22.349636144362\\
0.531506849315069	22.1930051779918\\
0.536986301369863	22.3769709257012\\
0.542465753424658	22.2207885913267\\
0.547945205479452	22.1185806482525\\
0.553424657534247	22.312294027459\\
0.558904109589041	22.4097851078674\\
0.564383561643836	22.612427348827\\
0.56986301369863	22.7288035851863\\
0.575342465753425	22.6353553467097\\
0.580821917808219	22.5166863383939\\
0.586301369863014	22.6986183043915\\
0.591780821917808	22.5426417716952\\
0.597260273972603	22.8145315711256\\
0.602739726027397	23.0005657588365\\
0.608219178082192	23.0694616736742\\
0.613698630136986	23.0702391426203\\
0.619178082191781	23.1583750960993\\
0.624657534246575	24.0164595028179\\
0.63013698630137	23.9869599933767\\
0.635616438356164	23.5797613023237\\
0.641095890410959	24.0636904916639\\
0.646575342465753	24.2243706705375\\
0.652054794520548	24.0513703978629\\
0.657534246575342	24.740494317547\\
0.663013698630137	24.8939988530451\\
0.668493150684932	24.9769756674199\\
0.673972602739726	24.7561783527261\\
0.679452054794521	24.45915212903\\
0.684931506849315	23.97773152534\\
0.69041095890411	23.8671289626198\\
0.695890410958904	23.5943768341869\\
0.701369863013699	23.7582529070878\\
0.706849315068493	23.6767449890338\\
0.712328767123288	24.1276720255388\\
0.717808219178082	24.5279535166927\\
0.723287671232877	24.5728731175713\\
0.728767123287671	24.495148879447\\
0.734246575342466	24.8099043747195\\
0.73972602739726	24.4887196997513\\
0.745205479452055	24.67266200416\\
0.750684931506849	24.4872510109992\\
0.756164383561644	24.3779743569918\\
0.761643835616438	24.1962367735837\\
0.767123287671233	24.3496483206278\\
0.772602739726027	24.1396474357997\\
0.778082191780822	23.8576082233113\\
0.783561643835616	24.0008016985818\\
0.789041095890411	23.8898149085748\\
0.794520547945205	23.9837080175619\\
0.8	23.891603474216\\
0.805479452054795	23.9616066801676\\
0.810958904109589	23.2919598192624\\
0.816438356164384	23.4066979385307\\
0.821917808219178	23.8681124036031\\
0.827397260273973	24.1375196303103\\
0.832876712328767	24.379121651556\\
0.838356164383562	24.3142511767261\\
0.843835616438356	24.3629692823121\\
0.849315068493151	24.428978781074\\
0.854794520547945	24.4551711336386\\
0.86027397260274	24.2290451495688\\
0.865753424657534	24.1349028405652\\
0.871232876712329	24.0885585614235\\
0.876712328767123	23.9616064581251\\
0.882191780821918	24.1802986331552\\
0.887671232876712	24.8561087333141\\
0.893150684931507	24.4839532939791\\
0.898630136986301	24.5189857778212\\
0.904109589041096	24.1241882947034\\
0.90958904109589	24.4693215068915\\
0.915068493150685	24.8537535844688\\
0.920547945205479	24.3862886780722\\
0.926027397260274	24.9126858260411\\
0.931506849315069	24.6062582081796\\
0.936986301369863	24.6687922904398\\
0.942465753424658	24.9733097598051\\
0.947945205479452	25.0150920742903\\
0.953424657534247	25.6689308942767\\
0.958904109589041	26.4928957542677\\
0.964383561643836	26.5368314985815\\
0.96986301369863	25.9292149064926\\
0.975342465753425	25.8178486463448\\
0.980821917808219	25.5611747406462\\
0.986301369863014	25.3340900195379\\
0.991780821917808	25.0049291634765\\
0.997260273972603	25.0271801691691\\
1.0027397260274	25.6277525930845\\
1.00821917808219	25.4141287552976\\
1.01369863013699	25.3317479940831\\
1.01917808219178	25.6721865632279\\
1.02465753424658	26.0353255604739\\
1.03013698630137	26.1846208739783\\
1.03561643835616	26.503506743192\\
1.04109589041096	26.7802955705328\\
1.04657534246575	26.6565948875242\\
1.05205479452055	26.79439083674\\
1.05753424657534	26.6973700494368\\
1.06301369863014	26.4074439783957\\
1.06849315068493	26.6078646802213\\
1.07397260273973	26.629463366574\\
1.07945205479452	26.5694574961471\\
1.08493150684932	26.6626201245868\\
1.09041095890411	26.9800828713803\\
1.0958904109589	27.1091493456968\\
1.1013698630137	27.173610920822\\
1.10684931506849	27.2667138853916\\
1.11232876712329	27.2838641759129\\
1.11780821917808	27.0242519674514\\
1.12328767123288	27.28345683838\\
1.12876712328767	27.755731063202\\
1.13424657534247	27.5710964809862\\
1.13972602739726	28.2296071572382\\
1.14520547945205	28.1668246922222\\
1.15068493150685	28.0803363826107\\
1.15616438356164	27.7635708096293\\
1.16164383561644	27.4895268078215\\
1.16712328767123	27.1657229941973\\
1.17260273972603	27.2834289697601\\
1.17808219178082	27.8187613665445\\
1.18356164383562	28.0018430921017\\
1.18904109589041	28.3018045179671\\
1.19452054794521	28.7819453943686\\
1.2	27.8609088957885\\
1.20547945205479	27.8025932392846\\
1.21095890410959	27.9249959820733\\
1.21643835616438	28.1756425791939\\
1.22191780821918	27.7123951618064\\
1.22739726027397	27.3656401262251\\
1.23287671232877	27.6321253711119\\
1.23835616438356	27.5921473093594\\
1.24383561643836	27.7813381223255\\
1.24931506849315	27.4487024500843\\
1.25479452054795	26.8950461214447\\
1.26027397260274	27.1436396641654\\
1.26575342465753	27.5399388685955\\
1.27123287671233	28.0963108699167\\
1.27671232876712	27.5550532864033\\
1.28219178082192	27.0988686977907\\
1.28767123287671	26.6108674919072\\
1.29315068493151	26.5975240513015\\
1.2986301369863	26.4005778162038\\
1.3041095890411	26.8153255586396\\
1.30958904109589	26.3427045666565\\
1.31506849315068	25.7950583810533\\
1.32054794520548	25.8570918727614\\
1.32602739726027	26.2191833244363\\
1.33150684931507	25.9690461478207\\
1.33698630136986	25.5817398881119\\
1.34246575342466	25.5373263395696\\
1.34794520547945	25.052235424149\\
1.35342465753425	25.0571899591069\\
1.35890410958904	25.2939159993374\\
1.36438356164384	25.3572831746782\\
1.36986301369863	24.7989605315734\\
1.37534246575342	24.6486330710395\\
1.38082191780822	24.5650625610929\\
1.38630136986301	25.0635051133033\\
1.39178082191781	25.3250683518287\\
1.3972602739726	25.3084169744438\\
1.4027397260274	25.689895277728\\
1.40821917808219	25.3766645478213\\
1.41369863013699	25.274670368261\\
1.41917808219178	25.6851568369779\\
1.42465753424658	26.1354405761691\\
1.43013698630137	26.3616471470753\\
1.43561643835616	26.5161044904102\\
1.44109589041096	26.3296191393876\\
1.44657534246575	26.5635359757152\\
1.45205479452055	26.2991729963892\\
1.45753424657534	26.4791982132571\\
1.46301369863014	26.5686395430197\\
1.46849315068493	26.9324571328576\\
1.47397260273973	26.7930229504549\\
1.47945205479452	26.9993329422386\\
1.48493150684932	27.2602068106497\\
1.49041095890411	26.9595287987316\\
1.4958904109589	27.0107791451428\\
1.5013698630137	27.5365171188423\\
1.50684931506849	27.5748395202495\\
1.51232876712329	27.4693894031075\\
1.51780821917808	27.6243502262938\\
1.52328767123288	27.5300520636004\\
1.52876712328767	27.6810234023789\\
1.53424657534247	27.634659445682\\
1.53972602739726	27.6283727590459\\
1.54520547945205	27.7863618122837\\
1.55068493150685	28.2582331763909\\
1.55616438356164	28.4194861362978\\
1.56164383561644	29.0136201604371\\
1.56712328767123	28.2557668490558\\
1.57260273972603	28.0954726498336\\
1.57808219178082	28.1788346388542\\
1.58356164383562	27.8822729359087\\
1.58904109589041	27.4375969252139\\
1.59452054794521	27.6465432055033\\
1.6	27.8568649860151\\
1.60547945205479	27.9575523181564\\
1.61095890410959	28.0959825944389\\
1.61643835616438	27.7884560381411\\
1.62191780821918	27.6160463196417\\
1.62739726027397	27.579501373952\\
1.63287671232877	27.3444146021897\\
1.63835616438356	27.4561124962768\\
1.64383561643836	28.2580198871244\\
1.64931506849315	28.0859365166433\\
1.65479452054795	28.3145033271557\\
1.66027397260274	27.9440727781634\\
1.66575342465753	28.412751862553\\
1.67123287671233	28.0635853612191\\
1.67671232876712	28.1371640436266\\
1.68219178082192	27.9180800620503\\
1.68767123287671	28.096926914437\\
1.69315068493151	28.1009380165114\\
1.6986301369863	28.0853941818126\\
1.7041095890411	29.1264684688065\\
1.70958904109589	28.8894865082158\\
1.71506849315069	28.8721002700503\\
1.72054794520548	29.8661223427195\\
1.72602739726027	29.4159777293389\\
1.73150684931507	29.6275348062667\\
1.73698630136986	29.2106639713721\\
1.74246575342466	29.6002498485806\\
1.74794520547945	29.7268910599289\\
1.75342465753425	29.9808727049468\\
1.75890410958904	29.8706924428993\\
1.76438356164384	29.9669696786033\\
1.76986301369863	30.5483169253338\\
1.77534246575342	29.6207369902145\\
1.78082191780822	28.9884730302452\\
1.78630136986301	29.1434719429293\\
1.79178082191781	28.8969437053142\\
1.7972602739726	28.7869874539724\\
1.8027397260274	28.2352300280935\\
1.80821917808219	27.9127990281627\\
1.81369863013699	27.7716659825658\\
1.81917808219178	27.5205048560789\\
1.82465753424658	27.2254921493829\\
1.83013698630137	27.0846293724796\\
1.83561643835616	26.7728036953596\\
1.84109589041096	27.207394795029\\
1.84657534246575	26.8777346188168\\
1.85205479452055	27.4707414126809\\
1.85753424657534	27.3436705321639\\
1.86301369863014	26.8550963741203\\
1.86849315068493	26.6535954764398\\
1.87397260273973	26.9537751106465\\
1.87945205479452	27.2999987594995\\
1.88493150684932	27.3537668447234\\
1.89041095890411	27.4505540425692\\
1.8958904109589	27.6650566522879\\
1.9013698630137	27.1636121271086\\
1.90684931506849	26.9701673970182\\
1.91232876712329	26.3694221760319\\
1.91780821917808	26.2279957781282\\
1.92328767123288	26.52394915003\\
1.92876712328767	26.7523395387064\\
1.93424657534247	27.4922507420172\\
1.93972602739726	27.54891200397\\
1.94520547945205	26.8133298335516\\
1.95068493150685	27.361989967768\\
1.95616438356164	27.7940551006636\\
1.96164383561644	27.5921470409598\\
1.96712328767123	27.6682949047439\\
1.97260273972603	27.9362333813047\\
1.97808219178082	28.0707665194469\\
1.98356164383562	28.3163683739342\\
1.98904109589041	28.2759999306577\\
1.99452054794521	28.2621302445958\\
2	28.2935725916915\\
};
\addplot [color=mycolor2, forget plot]
  table[row sep=crcr]{%
0	0\\
0.00547945205479452	0.0632570652411673\\
0.010958904109589	0.126334150936746\\
0.0164383561643836	0.188816013693762\\
0.0219178082191781	0.251112246800481\\
0.0273972602739726	0.313109249090495\\
0.0328767123287671	0.374843205455817\\
0.0383561643835616	0.436256191244343\\
0.0438356164383562	0.498450074963792\\
0.0493150684931507	0.561042410233937\\
0.0547945205479452	0.623673963561131\\
0.0602739726027397	0.685329738093333\\
0.0657534246575343	0.745938899338265\\
0.0712328767123288	0.805329995074487\\
0.0767123287671233	0.866277506569525\\
0.0821917808219178	0.927842233140023\\
0.0876712328767123	0.989574165494063\\
0.0931506849315069	1.05066772414052\\
0.0986301369863014	1.11102557857406\\
0.104109589041096	1.1717273152334\\
0.10958904109589	1.23342890964017\\
0.115068493150685	1.2940106906393\\
0.120547945205479	1.35430584735803\\
0.126027397260274	1.4145476575296\\
0.131506849315069	1.47353154766188\\
0.136986301369863	1.53302565716542\\
0.142465753424658	1.59226307747339\\
0.147945205479452	1.65268121403386\\
0.153424657534247	1.71284718785456\\
0.158904109589041	1.77327589859342\\
0.164383561643836	1.83296380487327\\
0.16986301369863	1.89225609391492\\
0.175342465753425	1.95081828465925\\
0.180821917808219	2.0096203754264\\
0.186301369863014	2.06922943479263\\
0.191780821917808	2.12845922107348\\
0.197260273972603	2.18796532261889\\
0.202739726027397	2.2471484022223\\
0.208219178082192	2.30639643418226\\
0.213698630136986	2.36641056663249\\
0.219178082191781	2.42650409989177\\
0.224657534246575	2.48591110391117\\
0.23013698630137	2.54476287398594\\
0.235616438356164	2.60350644333405\\
0.241095890410959	2.66212804442223\\
0.246575342465753	2.7202001284381\\
0.252054794520548	2.77838517318499\\
0.257534246575342	2.83736962067218\\
0.263013698630137	2.89619072489076\\
0.268493150684932	2.95487917520913\\
0.273972602739726	3.01214898416969\\
0.279452054794521	3.06894367824702\\
0.284931506849315	3.12489948147158\\
0.29041095890411	3.18062955792814\\
0.295890410958904	3.23666596724164\\
0.301369863013699	3.29360776495501\\
0.306849315068493	3.35166357029916\\
0.312328767123288	3.40964395006478\\
0.317808219178082	3.46741755618532\\
0.323287671232877	3.52571843945553\\
0.328767123287671	3.58434266822503\\
0.334246575342466	3.64347965511011\\
0.33972602739726	3.70314952349104\\
0.345205479452055	3.761788025952\\
0.350684931506849	3.82079240874877\\
0.356164383561644	3.87973398373708\\
0.361643835616438	3.93815325403598\\
0.367123287671233	3.99610317976631\\
0.372602739726027	4.05491246314907\\
0.378082191780822	4.11316245808758\\
0.383561643835616	4.17057302766533\\
0.389041095890411	4.22812600763876\\
0.394520547945205	4.287028672415\\
0.4	4.3459494341562\\
0.405479452054795	4.405088256806\\
0.410958904109589	4.46355823791796\\
0.416438356164384	4.52320482440171\\
0.421917808219178	4.58294297082831\\
0.427397260273973	4.64307182303066\\
0.432876712328767	4.70248958155059\\
0.438356164383562	4.76253504217622\\
0.443835616438356	4.82287322011392\\
0.449315068493151	4.88233322572765\\
0.454794520547945	4.94182289048158\\
0.46027397260274	5.0009766253266\\
0.465753424657534	5.06021146882764\\
0.471232876712329	5.11925968690202\\
0.476712328767123	5.1781668057724\\
0.482191780821918	5.23630775771579\\
0.487671232876712	5.29421378983262\\
0.493150684931507	5.35267620155936\\
0.498630136986301	5.41243484863589\\
0.504109589041096	5.47129643200637\\
0.50958904109589	5.53196162783887\\
0.515068493150685	5.59385233236235\\
0.520547945205479	5.65588362472748\\
0.526027397260274	5.71766222191621\\
0.531506849315069	5.77885662266094\\
0.536986301369863	5.84074008192276\\
0.542465753424658	5.90203663794511\\
0.547945205479452	5.96294772166957\\
0.553424657534247	6.02459261964713\\
0.558904109589041	6.08660831441141\\
0.564383561643836	6.14939750537037\\
0.56986301369863	6.21263268497533\\
0.575342465753425	6.27550866648697\\
0.580821917808219	6.33792669916683\\
0.586301369863014	6.40104997031027\\
0.591780821917808	6.46356658259499\\
0.597260273972603	6.52714530020956\\
0.602739726027397	6.59145364633057\\
0.608219178082192	6.65603338010743\\
0.613698630136986	6.72061636636248\\
0.619178082191781	6.78554923754219\\
0.624657534246575	6.85390054711261\\
0.63013698630137	6.92213405766537\\
0.635616438356164	6.98873260693758\\
0.641095890410959	7.0572824527397\\
0.646575342465753	7.12648294449186\\
0.652054794520548	7.19498045172378\\
0.657534246575342	7.26629060612067\\
0.663013698630137	7.33823000855958\\
0.668493150684932	7.41051104299317\\
0.673972602739726	7.48187998272077\\
0.679452054794521	7.55201677043188\\
0.684931506849315	7.62014803361207\\
0.69041095890411	7.68781677344411\\
0.695890410958904	7.75433980761867\\
0.701369863013699	7.82155445031417\\
0.706849315068493	7.88842396168183\\
0.712328767123288	7.9572123628763\\
0.717808219178082	8.02771144313598\\
0.723287671232877	8.09840351286431\\
0.728767123287671	8.16876080929321\\
0.734246575342466	8.24048080091028\\
0.73972602739726	8.31080466321524\\
0.745205479452055	8.3819319339424\\
0.750684931506849	8.45224628175316\\
0.756164383561644	8.52207949558574\\
0.761643835616438	8.59110884388078\\
0.767123287671233	8.66082021005682\\
0.772602739726027	8.72959436404646\\
0.778082191780822	8.79710407943251\\
0.783561643835616	8.86525901105868\\
0.789041095890411	8.93291200747838\\
0.794520547945205	9.00099200927351\\
0.8	9.06865170730575\\
0.805479452054795	9.13663275007479\\
0.810958904109589	9.20152825663336\\
0.816438356164384	9.26695518254066\\
0.821917808219178	9.33452837867523\\
0.827397260273973	9.40336068662751\\
0.832876712328767	9.47332748517067\\
0.838356164383562	9.54298855852971\\
0.843835616438356	9.61288077291595\\
0.849315068493151	9.68308756725131\\
0.854794520547945	9.75341994145172\\
0.86027397260274	9.82266546816308\\
0.865753424657534	9.89145647953943\\
0.871232876712329	9.96002279198606\\
0.876712328767123	10.027970110424\\
0.882191780821918	10.0969896956109\\
0.887671232876712	10.1693385968527\\
0.893150684931507	10.2398454549288\\
0.898630136986301	10.3105268776091\\
0.904109589041096	10.3792346225869\\
0.90958904109589	10.4496769615713\\
0.915068493150685	10.5220611354438\\
0.920547945205479	10.5920726999975\\
0.926027397260274	10.6647702273456\\
0.931506849315069	10.7358966299786\\
0.936986301369863	10.8073456616995\\
0.942465753424658	10.8803727908647\\
0.947945205479452	10.9536178299046\\
0.953424657534247	11.0302864051392\\
0.958904109589041	11.1112919434269\\
0.964383561643836	11.1925302470358\\
0.96986301369863	11.2705368330963\\
0.975342465753425	11.3479481781014\\
0.980821917808219	11.4239798506754\\
0.986301369863014	11.4987843343862\\
0.991780821917808	11.5718001898104\\
0.997260273972603	11.6449379497717\\
1.0027397260274	11.7213759230721\\
1.00821917808219	11.7966339510297\\
1.01369863013699	11.8714346184069\\
1.01917808219178	11.9481375362229\\
1.02465753424658	12.026880927885\\
1.03013698630137	12.1064681405312\\
1.03561643835616	12.1878675769308\\
1.04109589041096	12.2708490349677\\
1.04657534246575	12.3531198811908\\
1.05205479452055	12.436187593171\\
1.05753424657534	12.5186915492849\\
1.06301369863014	12.5995003141433\\
1.06849315068493	12.6814884051272\\
1.07397260273973	12.7636046462375\\
1.07945205479452	12.8453640411309\\
1.08493150684932	12.9276816648183\\
1.09041095890411	13.0119121410638\\
1.0958904109589	13.0969252226507\\
1.1013698630137	13.1823317335294\\
1.10684931506849	13.2683098100012\\
1.11232876712329	13.3543941158898\\
1.11780821917808	13.4388662076187\\
1.12328767123288	13.5249587724686\\
1.12876712328767	13.6140221310223\\
1.13424657534247	13.7019172234316\\
1.13972602739726	13.7940072454909\\
1.14520547945205	13.8856951665436\\
1.15068493150685	13.9768254423689\\
1.15616438356164	14.0658990581355\\
1.16164383561644	14.1531818111156\\
1.16712328767123	14.2383345297639\\
1.17260273972603	14.3242671910272\\
1.17808219178082	14.4137693470423\\
1.18356164383562	14.5045007080917\\
1.18904109589041	14.5972593607108\\
1.19452054794521	14.6932848913039\\
1.2	14.7830019183479\\
1.20547945205479	14.8723171398851\\
1.21095890410959	14.9624828426999\\
1.21643835616438	15.054401849949\\
1.22191780821918	15.1430587182135\\
1.22739726027397	15.2292565560245\\
1.23287671232877	15.3173584162794\\
1.23835616438356	15.4051730562309\\
1.24383561643836	15.4943591681649\\
1.24931506849315	15.5811175072817\\
1.25479452054795	15.6638049349026\\
1.26027397260274	15.7483343710067\\
1.26575342465753	15.8358219364197\\
1.27123287671233	15.927493532296\\
1.27671232876712	16.0150647700296\\
1.28219178082192	16.0991538121027\\
1.28767123287671	16.1794891023983\\
1.29315068493151	16.2597213831194\\
1.2986301369863	16.3384153213461\\
1.3041095890411	16.4203757309526\\
1.30958904109589	16.4985852755937\\
1.31506849315068	16.5724136846595\\
1.32054794520548	16.6467428592349\\
1.32602739726027	16.7240165999318\\
1.33150684931507	16.799240288207\\
1.33698630136986	16.8712632371208\\
1.34246575342466	16.9429164921848\\
1.34794520547945	17.0104934006935\\
1.35342465753425	17.0781127607128\\
1.35890410958904	17.147756089577\\
1.36438356164384	17.2179462072564\\
1.36986301369863	17.2832813017511\\
1.37534246575342	17.3472980617454\\
1.38082191780822	17.4105756494163\\
1.38630136986301	17.4783045676642\\
1.39178082191781	17.5483906403983\\
1.3972602739726	17.6183257945306\\
1.4027397260274	17.6917616137581\\
1.40821917808219	17.7622973020904\\
1.41369863013699	17.8318801462138\\
1.41917808219178	17.9053364266887\\
1.42465753424658	17.9830820996079\\
1.43013698630137	18.0630035732565\\
1.43561643835616	18.1444252957293\\
1.44109589041096	18.2240190058573\\
1.44657534246575	18.3059294910436\\
1.45205479452055	18.3851965255697\\
1.45753424657534	18.4662827121021\\
1.46301369863014	18.5482821748186\\
1.46849315068493	18.6340332931951\\
1.47397260273973	18.7183322972155\\
1.47945205479452	18.8048037599031\\
1.48493150684932	18.8940513304417\\
1.49041095890411	18.9800659090739\\
1.4958904109589	19.0666381399099\\
1.5013698630137	19.1589889606175\\
1.50684931506849	19.2517661579641\\
1.51232876712329	19.3433588902442\\
1.51780821917808	19.4367132917318\\
1.52328767123288	19.528984190275\\
1.52876712328767	19.6230113273272\\
1.53424657534247	19.7164934643969\\
1.53972602739726	19.8099012828121\\
1.54520547945205	19.9052133729604\\
1.55068493150685	20.0062813015576\\
1.55616438356164	20.1093408164915\\
1.56164383561644	20.2198285569515\\
1.56712328767123	20.3207224565006\\
1.57260273972603	20.4195615697916\\
1.57808219178082	20.5194840071605\\
1.58356164383562	20.6155043313587\\
1.58904109589041	20.7055953746619\\
1.59452054794521	20.7985109703985\\
1.6	20.8943086529232\\
1.60547945205479	20.9915055324662\\
1.61095890410959	21.0906529776994\\
1.61643835616438	21.1854071310537\\
1.62191780821918	21.2776628081138\\
1.62739726027397	21.3693815573084\\
1.63287671232877	21.4575916202754\\
1.63835616438356	21.5474949004212\\
1.64383561643836	21.6497375929933\\
1.64931506849315	21.7492916754965\\
1.65479452054795	21.8524749218981\\
1.66027397260274	21.9496831937512\\
1.66575342465753	22.0545764567355\\
1.67123287671233	22.1536500116556\\
1.67671232876712	22.2539714469876\\
1.68219178082192	22.3505156098957\\
1.68767123287671	22.4501985107396\\
1.69315068493151	22.549953655818\\
1.6986301369863	22.6494267524537\\
1.7041095890411	22.7681825307943\\
1.70958904109589	22.8824668881255\\
1.71506849315069	22.9964174621579\\
1.72054794520548	23.1298622623274\\
1.72602739726027	23.254303416575\\
1.73150684931507	23.3830633536546\\
1.73698630136986	23.5031377759043\\
1.74246575342466	23.6315031746919\\
1.74794520547945	23.7626227082173\\
1.75342465753425	23.899387796793\\
1.75890410958904	24.0336490444829\\
1.76438356164384	24.1701502965287\\
1.76986301369863	24.3204959683491\\
1.77534246575342	24.4482148485021\\
1.78082191780822	24.5601253056541\\
1.78630136986301	24.6760113837027\\
1.79178082191781	24.7854095190024\\
1.7972602739726	24.8918360632474\\
1.8027397260274	24.9829342547794\\
1.80821917808219	25.0648193434703\\
1.81369863013699	25.1425534767391\\
1.81917808219178	25.2126761322989\\
1.82465753424658	25.2735788489927\\
1.83013698630137	25.3299375607385\\
1.83561643835616	25.3759011010746\\
1.84109589041096	25.4368536339744\\
1.84657534246575	25.4860313825418\\
1.85205479452055	25.5571765729468\\
1.85753424657534	25.6234343238155\\
1.86301369863014	25.6701466429695\\
1.86849315068493	25.7084624265711\\
1.87397260273973	25.7598323078041\\
1.87945205479452	25.8269428765817\\
1.88493150684932	25.8966149440981\\
1.89041095890411	25.9711278581413\\
1.8958904109589	26.0569329226341\\
1.9013698630137	26.1148761229377\\
1.90684931506849	26.1614390712683\\
1.91232876712329	26.1704500209691\\
1.91780821917808	26.1700317108696\\
1.92328767123288	26.190757091113\\
1.92876712328767	26.2290545453532\\
1.93424657534247	26.3290220719051\\
1.93972602739726	26.4341421777383\\
1.94520547945205	26.4656927212098\\
1.95068493150685	26.5582162167422\\
1.95616438356164	26.7047574347523\\
1.96164383561644	26.8224506303438\\
1.96712328767123	26.9528379457174\\
1.97260273972603	27.1368224883473\\
1.97808219178082	27.3544465755923\\
1.98356164383562	27.6539521422602\\
1.98904109589041	27.9332709166904\\
1.99452054794521	28.198718479797\\
2	inf\\
};
\end{axis}
\end{tikzpicture}%

%% file: tikz/timestep1.tex
%
%
\definecolor{mycolor1}{rgb}{0.00000,0.44700,0.74100}%
\definecolor{mycolor2}{rgb}{0.85000,0.32500,0.09800}%
\begin{tikzpicture}

\begin{axis}[%
width=2.6in,
height=2.6in,
at={(0.758in,0.481in)},
scale only axis,
unbounded coords=jump,
xmin=0,
xmax=1,
xlabel style={font=\color{white!15!black}},
xlabel={Time (years)},
ymin=40,
ymax=240,
ylabel style={font=\color{white!15!black}},
ylabel={Value (\$)},
axis background/.style={fill=white},
legend style={legend cell align=left, align=left, draw=white!15!black}
]
\addplot [color=mycolor1]
  table[row sep=crcr]{%
0	150\\
0.01	144.572766560332\\
0.02	140.284374595137\\
0.03	143.099435466826\\
0.04	149.840372772691\\
0.05	152.091619242566\\
0.06	149.254573214412\\
0.07	155.340342676242\\
0.08	161.545311632303\\
0.09	162.516241618977\\
0.1	167.158092509725\\
0.11	170.608202193689\\
0.12	162.189012566074\\
0.13	168.348694403903\\
0.14	162.816745370715\\
0.15	157.90034329037\\
0.16	159.297341489868\\
0.17	156.848117227466\\
0.18	155.81432851911\\
0.19	160.452591328357\\
0.2	168.85094297482\\
0.21	167.179667885494\\
0.22	169.918724216261\\
0.23	167.894129834326\\
0.24	169.82152836432\\
0.25	173.557584198024\\
0.26	168.20449617842\\
0.27	170.515935342109\\
0.28	175.640809107227\\
0.29	176.45146126936\\
0.3	191.713113334718\\
0.31	185.879691216028\\
0.32	179.02946907748\\
0.33	163.365470918868\\
0.34	170.802541688242\\
0.35	178.207536883089\\
0.36	177.495661579932\\
0.37	185.469441087341\\
0.38	187.173009647669\\
0.39	189.894412883778\\
0.4	190.96679142013\\
0.41	195.167604314492\\
0.42	196.159665093779\\
0.43	223.497802333378\\
0.44	210.460448817993\\
0.45	190.947617958709\\
0.46	180.057100587455\\
0.47	170.213887953617\\
0.48	166.523571743326\\
0.49	165.120861450247\\
0.5	163.316638965878\\
0.51	167.737084995787\\
0.52	171.001803911789\\
0.53	177.424879488415\\
0.54	193.200221419979\\
0.55	205.015019210252\\
0.56	191.860680463391\\
0.57	169.51894055027\\
0.58	177.1636638866\\
0.59	160.903240147475\\
0.6	161.440310315097\\
0.61	161.726701275269\\
0.62	179.736328598652\\
0.63	179.114312803421\\
0.64	174.570871699034\\
0.65	176.629146703869\\
0.66	178.799961766153\\
0.67	179.426165804434\\
0.68	173.966402429087\\
0.69	163.331893829852\\
0.7	165.916623996198\\
0.71	154.776407485765\\
0.72	146.788518253838\\
0.73	156.558878616071\\
0.74	153.279727892943\\
0.75	152.20430412423\\
0.76	159.052145694492\\
0.77	156.665480725007\\
0.78	164.72878443143\\
0.79	161.886669529183\\
0.8	170.084625564108\\
0.81	175.436632425738\\
0.82	173.568099988016\\
0.83	166.055228139172\\
0.84	157.394548791548\\
0.85	155.269180846077\\
0.86	151.86768837712\\
0.87	148.764467202759\\
0.88	156.080296347018\\
0.89	153.757063424085\\
0.9	162.549498965095\\
0.91	158.228769778068\\
0.92	165.923163711624\\
0.93	161.590491075621\\
0.94	163.017155703087\\
0.95	170.929501649551\\
0.96	167.391497916647\\
0.97	163.723359989896\\
0.98	180.123452099025\\
0.99	188.688263702151\\
1	184.612560942868\\
};
\addlegendentry{Stock Price}

\addplot [color=mycolor2]
  table[row sep=crcr]{%
0	50\\
0.01	50.9392868837851\\
0.02	51.8349362974149\\
0.03	52.7596509361226\\
0.04	53.7545887007044\\
0.05	54.7732783447918\\
0.06	55.7618932679485\\
0.07	56.8159696426192\\
0.08	57.9375171545725\\
0.09	59.0698092669379\\
0.1	60.2537242414091\\
0.11	61.4764641200942\\
0.12	62.6036828349\\
0.13	63.8017490467772\\
0.14	64.9356190827505\\
0.15	66.0117734451391\\
0.16	67.1046456006247\\
0.17	68.168119019687\\
0.18	69.2190886301901\\
0.19	70.3273989995731\\
0.2	71.5407548183417\\
0.21	72.7330645654282\\
0.22	73.9605860352887\\
0.23	75.1619264259649\\
0.24	76.3887308438406\\
0.25	77.6654518858786\\
0.26	78.8699539774534\\
0.27	80.1062307022607\\
0.28	81.4137984611854\\
0.29	82.7329009830722\\
0.3	84.2701515341923\\
0.31	85.722976213243\\
0.32	87.0751766566702\\
0.33	88.1936827429287\\
0.34	89.4250161607895\\
0.35	90.7704225687583\\
0.36	92.1048342188709\\
0.37	93.5659777734591\\
0.38	95.0547406042639\\
0.39	96.5882671853924\\
0.4	98.1398118382454\\
0.41	99.762724002416\\
0.42	101.402891040579\\
0.43	103.523022196424\\
0.44	105.410388341726\\
0.45	106.942959292718\\
0.46	108.273907078591\\
0.47	109.419192083724\\
0.48	110.493627910331\\
0.49	111.54070288191\\
0.5	112.551832362667\\
0.51	113.653393397289\\
0.52	114.823177097046\\
0.53	116.129878904158\\
0.54	117.77992762497\\
0.55	119.692884551711\\
0.56	121.306831653335\\
0.57	122.400984997256\\
0.58	123.677473244202\\
0.59	124.557227834732\\
0.6	125.450599502981\\
0.61	126.351501312234\\
0.62	127.726927290125\\
0.63	129.085711300371\\
0.64	130.318362251688\\
0.65	131.610062023829\\
0.66	132.965857747629\\
0.67	134.340838642375\\
0.68	135.545232492562\\
0.69	136.406440160883\\
0.7	137.354086748392\\
0.71	137.917402095765\\
0.72	138.195345478334\\
0.73	138.835724179519\\
0.74	139.350050670393\\
0.75	139.821515682903\\
0.76	140.578811881002\\
0.77	141.232431063107\\
0.78	142.253164883438\\
0.79	143.13861657983\\
0.8	144.434624975416\\
0.81	146.01283971996\\
0.82	147.487340803078\\
0.83	148.519620678973\\
0.84	149.010205318592\\
0.85	149.359150865296\\
0.86	149.46506745228\\
0.87	149.332213018157\\
0.88	149.809936978799\\
0.89	150.076426217638\\
0.9	151.223409719404\\
0.91	151.889962862957\\
0.92	153.519674030824\\
0.93	154.529917131199\\
0.94	155.778443639934\\
0.95	158.611554255633\\
0.96	160.559333701109\\
0.97	161.283166218153\\
0.98	170.217115117718\\
0.99	187.726402618118\\
1	-inf\\
};
\addlegendentry{Accumulated Cash}

\end{axis}
\end{tikzpicture}%

%% file: tikz/time.tex
%
%
\definecolor{mycolor1}{rgb}{0.00000,0.44700,0.74100}%
\begin{tikzpicture}

\begin{axis}[%
width=4.521in,
height=2.7in,
at={(0.758in,0.481in)},
scale only axis,
xmin=0,
xmax=10000,
xlabel style={font=\color{white!15!black}},
xlabel={Number of Time Steps (N)},
ymin=0,
ymax=3,
ylabel style={font=\color{white!15!black}},
ylabel={Time Perfomance (s)},
axis background/.style={fill=white}
]
\addplot [color=mycolor1, forget plot]
  table[row sep=crcr]{%
5	0.429\\
10	0.439\\
50	0.43\\
100	0.4672\\
500	0.5286\\
1000	0.6642\\
5000	1.6248\\
10000	2.815\\
};
\end{axis}
\end{tikzpicture}%

%% file: tikz/dime.tex
%
%
\definecolor{mycolor1}{rgb}{0.00000,0.44700,0.74100}%
\begin{tikzpicture}

\begin{axis}[%
width=4.521in,
height=2.7in,
at={(0.758in,0.481in)},
scale only axis,
xmin=0,
xmax=10000,
xlabel style={font=\color{white!15!black}},
xlabel={Number of Different Stocks (m)},
ymin=0.3,
ymax=0.8,
ylabel style={font=\color{white!15!black}},
ylabel={Time Perfomance (s)},
axis background/.style={fill=white}
]
\addplot [color=mycolor1, forget plot]
  table[row sep=crcr]{%
1	0.5418\\
1000	0.5406\\
3000	0.581\\
5000	0.6302\\
6000	0.648\\
7000	0.6676\\
8000	0.666\\
9000	0.7034\\
10000	0.7316\\
};
\end{axis}
\end{tikzpicture}%

%% file: sections/discussion.tex
\section{Discussion}
This section will present the main discussion points that arise when analyzing the model. In \autoref{sec:error}, a modeling error due to discretization will be discussed, and \autoref{sec:error2} discusses an assumption regarding interest rates. 

\subsection{Error at Last Time Step due to Discretization}
\label{sec:error}
\begin{figure}
    \centering
    \input{tikz/error.tex}
    \caption{Rendering of model at $\Delta t = 0.05$}
    \label{fig:error}
\end{figure}
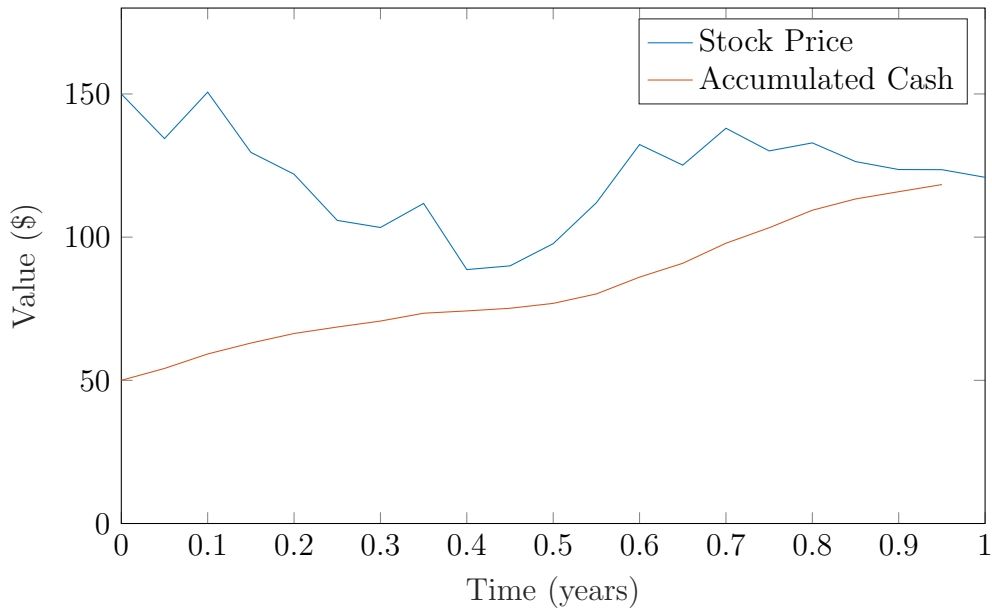

When analyzing the plots generated by the model, it should be noted that the very last time step is not plotted. This can clearly be seen in \autoref{fig:error}. The reason for this can be found in \autoref{eq:rr}, which is being used to calculate $R(s)$. At the terminal time $T$, $R(T)$ will have to be 0. Subsequently, \autoref{eq:ur} makes use of $R(s)^{-1}$. With $R(T)=0$, this automatically means that $R(T)^{-1} \rightarrow \infty$. Therefore, also $u(T) \rightarrow \infty$. As this result can't be right, this very last time step  will not be considered and will be excluded from the plot. 

This result seems contradictory to the method, which is designed to ensure that at the terminal time $x(T)=S(T)$. However, this turns out only to be true when the equations are evaluated continuously in time. The singularities that arise at the terminal time are defined, such that the desired solution can be obtained. When making use of the discretized model, these singularities are not defined, and therefore, the observed error is obtained. \\

The impact of this error, however, is not significant. Whereas for relatively large time steps, such as in \autoref{fig:error}, this error can be in the order of several dollars, for small time steps such as the ones used in Section 3, the order of error might be several cents.

\subsection{Assumption on Constant Interest Rate}
\label{sec:error2}
The next discussion point will deal with the assumption in the model that the interest rate $r$ is constant over the entire time interval in which the cash is accumulated. However, in reality 
this is not the case. \\

Banks can change their interest rates on a daily basis. Even though the daily changes in the interest rates may be relatively small, this could still mean that at the terminal time, which may be in the course of years, the interest is significantly different compared to the time at which the cash accumulation started. 
\\

Therefore, in future research, a way has to be found to incorporate this changing interest rate into the model. A suggestion to model the interest rates, is to model it with its own Wiener Process,  like the stock prices are modeled. However, it should be verified if the equations still produce the right outcome. If not, further research must be conducted such that the equations are found to produce the right income with changing interest rates.

%% file: tikz/error.tex
%
%
\definecolor{mycolor1}{rgb}{0.00000,0.44700,0.74100}%
\definecolor{mycolor2}{rgb}{0.85000,0.32500,0.09800}%
\begin{tikzpicture}

\begin{axis}[%
width=4.521in,
height=2.7in,
at={(0.758in,0.481in)},
scale only axis,
unbounded coords=jump,
xmin=0,
xmax=1,
xlabel style={font=\color{white!15!black}},
xlabel={Time (years)},
ymin=0,
ymax=180,
ylabel style={font=\color{white!15!black}},
ylabel={Value (\$)},
axis background/.style={fill=white},
legend style={legend cell align=left, align=left, draw=white!15!black}
]
\addplot [color=mycolor1]
  table[row sep=crcr]{%
0	150\\
0.05	134.412563000952\\
0.1	150.624360138697\\
0.15	129.604946703725\\
0.2	121.971660933281\\
0.25	105.844284342567\\
0.3	103.369347157754\\
0.35	111.771041253269\\
0.4	88.6684273434984\\
0.45	89.9737209797482\\
0.5	97.7214349940882\\
0.55	112.011585482159\\
0.6	132.337420425953\\
0.65	125.142644521665\\
0.7	138.025047067435\\
0.75	130.110436896305\\
0.8	132.91513933753\\
0.85	126.388184106749\\
0.9	123.608602248707\\
0.95	123.55608125729\\
1	120.890317709713\\
};
\addlegendentry{Stock Price}

\addplot [color=mycolor2]
  table[row sep=crcr]{%
0	50\\
0.05	54.1473998898559\\
0.1	59.2010656482149\\
0.15	63.0151399175516\\
0.2	66.3522677048167\\
0.25	68.6116142193972\\
0.3	70.695794630013\\
0.35	73.4320348111494\\
0.4	74.235787084381\\
0.45	75.1615024065607\\
0.5	76.8688900905466\\
0.55	80.1756189180588\\
0.6	86.0402723976771\\
0.65	90.8743166550957\\
0.7	97.8711541289327\\
0.75	103.279305700928\\
0.8	109.396502936093\\
0.85	113.329011750708\\
0.9	115.867479236196\\
0.95	118.357281215834\\
1	-inf\\
};
\addlegendentry{Accumulated Cash}

\end{axis}
\end{tikzpicture}%

%% file: sections/conclusions.tex
\section{Conclusions}

The aim of this report was to create a model to solve the problem of cash accumulation strategies for purposes of buying things with an unknown future price, like assets, at a predefined terminal time. The model is based on an optimal replication of random claims with ordinary integrals. For the model the stock prices were modeled after a discretized Wiener Process. Subsequently, all other provided equations for this problem were discretized in order to be simulated in the model.\\

The model is able to solve three different problems regarding buying assets. The first is to accumulate cash to buy one or a multiple of the same asset. The second is accumulating cash to buy a set consisting of different, independently behaving assets. Finally, the model can accumulate a proportion of the excess achieved by a certain asset. \\

One drawback of the model consists of the fact that due to discretization the last time step cannot be modeled correctly. However, this is not of great significance when a large number of time steps is used, and it only leads to minor inaccuracies in the end result. Further research could be conducted in correcting the assumption that interest rates are constant over the entire interval. 

%% file: references.tex
\section*{References}

\begin{hangparas}{.25in}{1} 
Brunick, G., \& Shreve, S. (2013). MIMICKING AN ITO PROCESS BY A SOLUTION OF A STOCHASTIC DIFFERENTIAL EQUATION. \textit{The Annals of Applied Probability}, 23(4), 1584-1628\\
\end{hangparas}

\begin{hangparas}{.25in}{1} 
Dokuchaev, N. (2013). Optimal replication of random vectors by ordinary integrals. \textit{Systems \& Control Letters}, 62(1), 43-47.\\
\end{hangparas}

\begin{hangparas}{.25in}{1} 
Dokuchaev, N. (2013). Optimal replication of random claims by ordinary integrals with applications in finance. \textit{SIAM Conference on Control and Its Applications} (pp. 59-66). San Diego: Society for Industrial and Applied Mathematics.\\
\end{hangparas}

\begin{hangparas}{.25in}{1}
Higham, D. J. (2001). An Algorithmic Introduction to Numerical Simulation of Stochastic Differential Equations. \textit{SIAM Review}, 43(3), 525-546.\\
\end{hangparas}

\begin{hangparas}{.25in}{1}
Haug, E. G. (2007). \textit{The Complete Guide to Option Pricing Formulas}. New York: McGraw-Hill Education.\\
\end{hangparas}

\begin{hangparas}{.25in}{1}
Huijskens, T. (2013). \textit{A numerical method for backward stochastic differential equations with applications in finance}. Delft: TU Delft \\
\end{hangparas}

\begin{hangparas}{.25in}{1} 
Hull, J. C. (2012). \textit{Options, Futures and Other Derivatives} (8th ed.). Toronto: Prentice Hall.\\
\end{hangparas}